\documentclass[journal ]{new-aiaa}
\usepackage[utf8]{inputenc}
\usepackage{textcomp}

\usepackage{graphicx}
\usepackage{amsmath}
\usepackage[version=4]{mhchem}
\usepackage{siunitx}
\usepackage{longtable,tabularx}
\usepackage{mathdots}
\usepackage{tikz}

\usepackage[nottoc]{tocbibind}
\usepackage{comment}
\newcommand{\sign}{\text{sgn}}

\title{Cooperative Aerial Transportation of Nonuniform Load through Quadrotors by Elastic and Flexible Cables}

\author{Ali Akbar Rezaei Lori\footnote{Graduate Student, Department of  Mechanical Engineering, Isfahan University of Technology; \href{mailto:aa.rezaei@alumni.iut.ac.ir}{aa.rezaei@alumni.iut.ac.ir} (Corresponding
Author).},  Mohammad Danesh\footnote{Associate Professor, Department of  Mechanical Engineering, Isfahan University of Technology.}, and Iman Izadi\footnote{Associate Professor, Department of Electrical and Computer Engineering, Isfahan University of Technology.}}
\affil{Isfahan University of Technology, Isfahan 84156-83111, Iran}

\begin{document}

\maketitle

\begin{abstract}

In this paper, first the full dynamics of aerial transportation of a rigid body with arbitrary number of quadrotors is derived. Then a control strategy is proposed to convey the nonuniform rigid body appropriately to the desired trajectory. In the dynamical model of this transportation system, not only the load is considered as a nonuniform and non-homogeneous rigid body but also  mass, flexibility, and tension of the cables are considered. Each cable is modeled as successive masses, springs, and dampers where each mass, spring, and damper has 4 degrees of freedom (DOF). The Euler-Lagrange equations are used to derive the motion equation. The control strategy includes three loops of attitude control, formation control, and navigation control. The sliding mode control is designed based on multi-agent systems for the formation control where the controller is proven to be  asymptotically stable. The navigation control loop, based on the load states, guarantees that the load reaches the desired location. Finally, numerical examples and simulations are presented to verify the appropriate operation of the proposed system for transporting both homogeneous and non-homogeneous bodies by spreading quadrotors according to mass distribution of the body.  
\end{abstract}

\begin{table}
\caption{\label{Notations}Dynamic notations}\centering
\begin{tabular}{lcc} \hline 
Parameter  & Value  \\ \hline 
$\{X,Y,Z\}\in\mathbb{R}^3 $ & Inertia frame\\
$\{x,y,z\} \in\mathbb{R}^3$ & Body frame attached to the load\\
$m_l\in\mathbb{R}$ & mass of payload \\   
$J_l\in\mathbb{R}^{3\times3}$ &  moment inertia of payload \\ $r_{l}\in\mathbb{R}^3$ & position of payload\\   
$v_{l}\in\mathbb{R}^3$ & velocity of payload\\
$\boldsymbol{\delta F_{r_l}}\in \mathbb{R}^{3}$ & the disturbances and unmodeled dynamics effects on the load position \\ 
${\eta }\in\mathbb{R}^3$ & load's angles\\   
$\Omega_l\in\mathbb{R}^3$ & angular velocity of the load \\  
$\boldsymbol{\delta F_{\eta_l}}\in \mathbb{R}^{3}$& the disturbances and unmodeled dynamics effects on load's angles\\
$m_{ij}\in\mathbb{R}$ & mass of at the end of component $j$th in cable $i$th  \\   
$K_{ij}\in\mathbb{R}$ & elasticity coefficient of  $j$th spring of cable $i$th \\   
$b_{ij}\in\mathbb{R}$ & viscosity coefficient of  $j$th damper the $i$th cable \\   
$L_{ij}\in\mathbb{R}$ & rest length of the  $j$th component of the $i$th cable \\ $q_{ij}\in\mathbb{R}^3$ & the direction of the $j$th component of the $i$th cable\\
$\omega_{ij}\in\mathbb{R}^3$ & angular velocity of the $j$th component of the $i$th cable\\ $\delta \boldsymbol{F_{q_{ij}}}\in \mathbb{R}^3$ & the disturbances and unmodeled dynamics effects on the $q_{ij}$\\ $\delta F_{l_{ij}}\in \mathbb{R}$ &the disturbances and unmodeled dynamics effects on the $l_{ij}$\\ 
$\{x_i,y_i,z_i\} \in\mathbb{R}^3$ & Body frame attached to the $i$th quadrotor\\
$J_i\in\mathbb{R}^{3\times3}$ &  moment inertia of $i$th quadrotor \\   
$d_i\in\mathbb{R}^3$ &  connection of $i$th quadrotor \\
$m_i\in\mathbb{R}$ & mass of $i$th quadrotor \\     
$\Omega_{i}\in\mathbb{R}^3$ & angular velocity of the $i$th quadrotor \\   
$r_i\in\mathbb{R}^3$ & the $i$th quadrotor position \\   
$v_{i}\in\mathbb{R}^3$ & the $i$th quadrotor velocity \\ $f_{i}\in\mathbb{R}$ & the generated trust by the $i$th quadrotor \\$\boldsymbol{\delta T_i}\in \mathbb{R}^{3}$ &the disturbances and unmodeled dynamics effects on the rotation of the $i$th quadrotor\\
$\boldsymbol{\tau}_{i}\in\mathbb{R}^3$ & the produce torque by the $i$th quadrotor \\\hline 
\end{tabular}
\end{table}

\section{Introduction}
In recent years, due to some challenges such as traffic jams, and difficult accessibility of remote or dangerous regions,  aerial transportation of rigid bodies through aerial vehicles has been an active field of research. Many researchers have focused on quadrotors as one of the most popular unmanned aerial vehicles (UAVs). Broad applications of aerial load transportation include search and rescue, emergency response, package delivery, construction, and firefighting. 

By and large, carrying a rigid body with quadrotors has been done by grasping the load (~\cite{1pp},~\cite{2pp},~\cite{3pp} and~\cite{4pp}) or by hovering the load from a cable. Load grasping by quadrotors has some limitations like landing in bumpy and narrow areas. Therefore, load transportation has been usually investigated by using cables. There are some methods for modeling and control of a cable-suspended load with a quadrotor.

In the primary works, due to the complexity of the dynamics of aerial load transportation, it was simplified by considering some assumptions such as the motion equations were supposed to be decoupled, and external force and momentum were assumed instead of load effects~\cite{5pp}. Some works like~\cite{6pp} and~\cite{7npp} obtained motion equations of aerial load transportation of a slung load in 2D coordinates where the sliding mode and feedback linearization control were used for transitional and rotational movements, respectively. Then, in the next works, the dynamics of a quadrotor and a suspended rigid body were modeled together using different methods, and different control approaches were applied to control the system. In these researches, the load was considered a mass point and the cable was assumed massless, rigid, and inelastic. For instance, motion equations of a quadrotor with a slung load were obtained by the Lagrange method in~\cite{7pp}, and an anti-swing control algorithm was designed to reduce the load fluctuations. In~\cite{8pp}, a feedback linearization controller and a neural network controller were designed for position and attitude control of an aerial transportation system, where the Udwadia-Kalaba equations were employed to derive its dynamics. In~\cite{9pp}, an interconnection and damping assignment-passivity-based control (IDA-PBC) strategy was utilized to reduce swings of a point mass which was carried through a quadrotor.~\cite{10pp} used an optimal control approach for path planning of the maximum mass of the load transportation with minimum swings. In the study, the optimal cable length was also determined for the maximum load mass. In~\cite{11pp} a nonlinear dynamic model of aerial load transportation was presented using Kane's equation and a backstepping controller was designed to dampen the load swings and compensate for wind effects. In~\cite{12pp}, the mathematical model of a quadrotor with a point mass slung load was obtained as coupled and parallel subsystems; the cable-load subsystem and the quadrotor subsystem. Afterwards, a controller which was based on the trajectory tracking of the quadrotor was presented. In another study where a quadrotor carried a point in dense environments was investigated\cite{33pp}. In \cite{13pp}, the control of an aerial transportation system of a cable-suspended payload was designed where the controller had a fast part for the load swings and a slow part for trajectory control. Since adaptive sliding mode control approaches are common for systems in environments with unknowns and chaotic behaviors \cite{32pp}, some works use that method. For example, \cite{14pp} investigated the aerial transporting of a mass point that was connected to a quadrotor through four cables. The dynamics of load and quadrotor were more coupled since the load moments have impacts on the rotational movements of the quadrotor. An adaptive sliding mode control approach was presented for carrying a varying load by a quadrotor where the mass of the load has consisted of a static part and an exponentially decreasing part \cite{15pp}. A robust control was applied in \cite{15npp} to carry an unknown payload as well as manage the disturbances caused by wind forces and load swings.

In some other robotic systems that use cable, like the cable parallel robots (\cite{gibbs}, \cite{zare1}, and \cite{zare2}), the dynamics of cable is also considered which they use different methods like Gibbs-Appell or Lagrangian to model the cable. Therefore, the dynamics of cable can be effective in carrying a payload and it is required that the physical features of the cable are considered in the motion equations. In some studies such as~\cite{16pp}, the cable dynamics was considered in the motion equations and several serially connected rigid links were assumed to model the cable. In another study~\cite{17pp}, an approach for lifting a suspended payload was presented according to the relation between the load and quadrotor distance and the length of cable. This methodology divided the lifting process into three steps: setup, pull, and raise, where if the distance between the load and quadrotor was less than the length of cable, the impact of the load was ignored. A parallel spring and damper was utilized to indicate the effect of the cable elasticity on aerial transporting of a point mass~\cite{18pp} and \cite{elastic1}. Some works like~\cite{19pp} and~\cite{20pp} applied several masses, springs, and dampers to show the impact of the mass, flexibility, and elasticity of the cable in the motion equations.       

For the sake of resolving restrictions like the limited capacity of a quadrotor for transportation of a load or oscillation of the load, several quadrotors are employed simultaneously and different approaches have been used to control the system. Multiple quadrotors were taken to convey a point mass load which was hanged via massless elements~\cite{21pp}, where the motion equations of the system were derived by Hamilton’s principle. PD and PID control are simultaneously used in attitude control of a transportation system to damp the swings of a mass point during transporting the load hanged by massless rigid elements from several quadrotors \cite{21npp}. In~\cite{22pp}, the motion equation of the load and quadrotors were decoupled by assuming that the load mass is shared between quadrotors equally. In that research, an adaptive force control which is based on a consensus algorithm was designed to control the system. The mathematical model of a system that consists of two quadrotors and a cable-suspended point mass load was obtained by the Euler-Lagrange method and carrying the payload was done along with collision avoidance of obstacles that was performed by potential field~\cite{23pp}. LQR formation control was built for a group of quadrotors that carry a swing point mass where the transportation system was modeled via the Udwadia-Kalaba approach~\cite{24pp}. Some works employed continuum deformation of multiple quadrotors to deliver a slung load~\cite{25pp} and~\cite{27pp}. In these works, the group motion of quadrotors was assumed as a continuum deformation and distances between agents can increase and decrease homogeneously.

 Since the payload is not a point mass in reality, some works considered it as a rigid body. In \cite{27npp}, the dynamics of the system was decoupled by considering the impact of each cable force on the corresponding quadrotor. Then, the system is controlled by the PID method where the load was taken as a leader .~\cite{28pp} studied the aerial transportation of a 6-DOF rigid body that was suspended from several quadrotors via massless rigid components and designed a geometry control to track the desired position of the load. An observer system is designed by \cite{28npp} to estimate not only the load mass and inertial but also the cable force. Moreover, different prediction methods such as the Kalman filter, recursive least squares, and maximum likelihood estimation were compared.  In~\cite{29pp}, the motion equations of aerial transportation of a fluid container by using three quadrotors were given by Euler-Lagrange equations in 2D coordinates. Multiple masses and springs were used to model the fluid dynamics, and cables were assumed as massless rigid links. The LQR approach was used to control the linear motion equations of aerial load transportation in~\cite{30pp}. In~\cite{31pp}, a robust control was designed for carrying a platform through three quadrotors where two massless cables were connected to each quadrotor. 

 In this paper, firstly unlike previous works, the mass, flexibility (bending) and elasticity (increasing the length) of the cables are not neglected, in addition the load is assumed as a non-uniform 6-DOF rigid body (inertia moment is not ignored), and complete coupled dynamics of the aerial transportation with an arbitrary number of quadrotors is presented. Secondly, a hierarchical and decentralized control strategy is proposed to appropriately convey the rigid body to the desired location. Unlike previous research, this control strategy is used for full dynamics of quadrotors, cables, and the rigid body; where each cable is modeled as successive masses, springs, and dampers with 4 DOF and it is shown that these components behave like a cable. Generally, one way to navigate the multi-agent systems is that the system follows a leader, which is done in the target tracking~\cite{arm1}  or autonomous cars \cite{Mohammad1} and \cite{mohammad2}, or using image processing-based approaches for each agent \cite{pedram2}, meanwhile, in the aerial load transportation system in here, the agents (quadrotors) have interaction with each other, therefore choosing an appropriate leader is crucial. Thus, the proposed control strategy is based on the load states (load-leading based), in other words, there is a feedback from the load states that helps to guide the load on the desired trajectory (with the desired position and velocity). Furthermore, the capability of this strategy for transporting homogeneous as well as nonhomogeneous bodies is shown for a proper spreading of quadrotors. It is worth mentioning that the control system includes three parts; attitude control, formation control, and navigation control. Asymptotically stability is proven for the designed sliding mode formation control system. 

In section 3, motion equations are presented for the above mentioned aerial body carrying system based on Euler-Lagrange equations. Then in section 4, a control law is proposed for the transportation system to convey the body on the desired trajectory. The proposed dynamic model and control strategy of the transportation system is confirmed by numerical simulations for homogeneous and nonhomogeneous bodies in section 5. The paper is concluded in section 6.  

\section{Preliminaries}
\begin{figure} 
\centering 
\includegraphics*[width=3.41in]{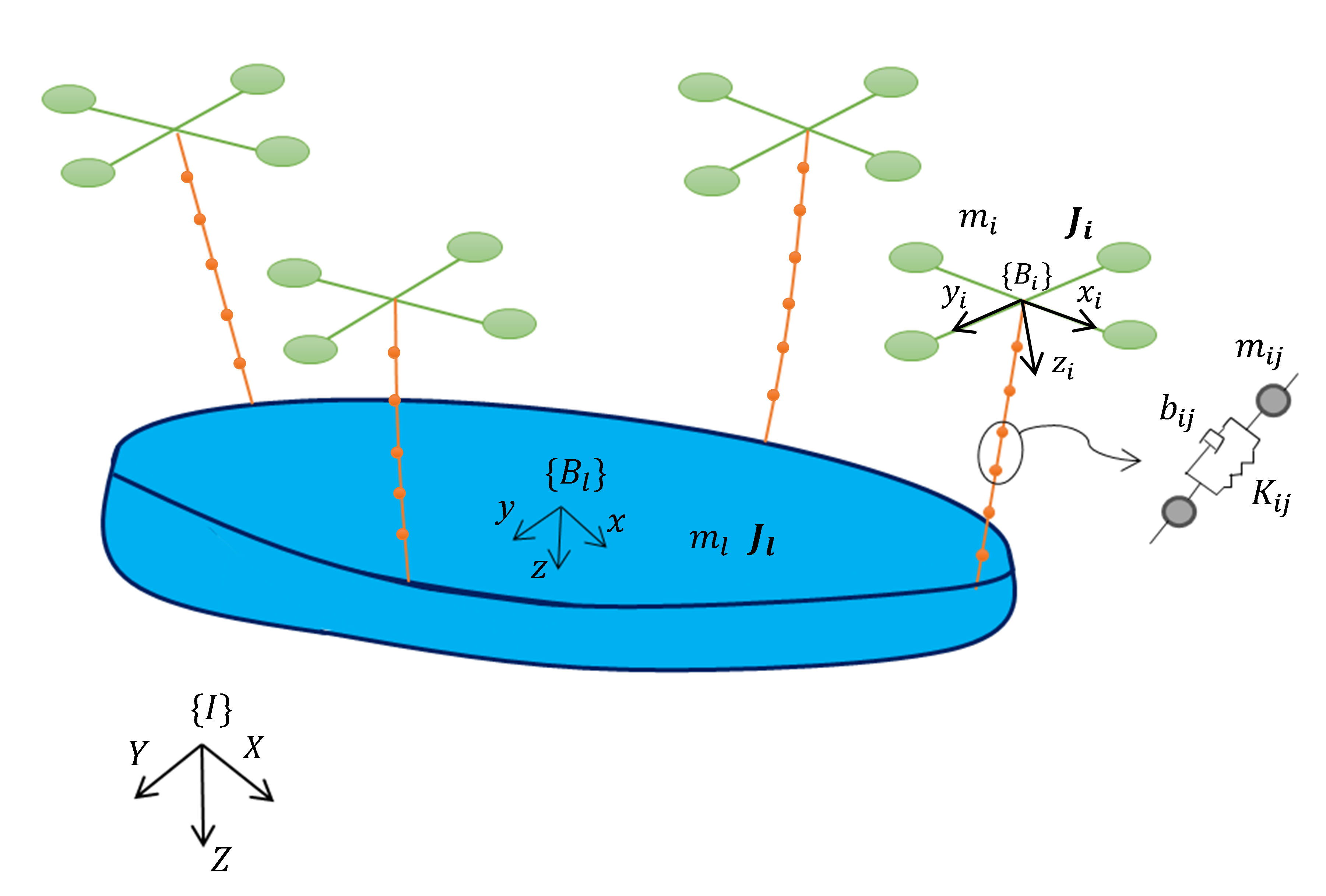}
\caption{Model of the aerial transportation system including the quadrotors, suspended rigid body, and connecting cables}
\label{fig1}
 \end{figure}

Consider $n$ quadrotors with masses ${m}_{i}\in \mathbb{R}$ and symmetric moments of inertia $\boldsymbol{J_i}\in \mathbb{R}^{3\times 3}$, $i=1, \cdots, n$. A suspended body with mass $m_l \in \mathbb{R}$ and moment of inertia $\boldsymbol{J_l}   \in \mathbb{R}  ^{3\times 3}$ is connected to them through cables. The cables are modeled as serial connections of several parallel springs and dampers with the point masses $m_{ij}   \in \mathbb{R}$, $i=1, \cdots, n$, at the end of the $j$th spring and damper of the $i$th cable as depicted in Figure.~\ref{fig1} (each cable can have ${n}_{i}$ different number of springs, dampers and masses). 

Three coordinates are used for describing the locations of quadrotors, load and cables point masses. First is the inertial reference frame $  \{X,Y,Z\}=\{ \boldsymbol{e_1}, \boldsymbol{e_2}, \boldsymbol{e_3}\}$ with the axis $ \boldsymbol{e_3}={\left[0,\ 0,\ 1\right]}^T   \in \mathbb{R}^3$ pointing downwards alongside the direction of the gravity and the other axes form an orthonormal frame. The other coordinates are the body-fixed frames $\{x,y,z\}$ and $\{x_{i},y_{i}, z_{i}\}$ that are attached to the load and the $i$th quadrotor, respectively. The location of the load mass center in the inertial reference frame is denoted by $\boldsymbol{r_l}\in \mathbb{R}^3$ and its attitude is given by $\boldsymbol{R_l}\in  \text{SO}(3)$, where $\text{SO}(3)=\left\{\boldsymbol{R}\in \mathbb{R}^{3\times 3}\ \right|\ {\boldsymbol{R}}^T\boldsymbol{R} = \boldsymbol{I},\det\left[\boldsymbol{R}\right]=\ 1\}$ determines the special orthogonal group, and $\boldsymbol{R}\ $is defined by 
\begin{equation}\label{GrindEQ__1}
\boldsymbol{R}=\left[ \begin{array}{ccc}
\cos\theta \cos\psi  & \cos\psi \sin\theta \sin\varphi -\sin\psi \cos\varphi  & \sin\psi \sin\varphi +\cos\psi \cos\varphi \sin\theta  \\ 
\sin\psi \cos\theta  & \cos\varphi \cos\psi +\sin\varphi \sin\theta \sin\psi  & \sin\psi \cos\varphi \sin\theta -\cos\psi \sin\varphi  \\ 
-\sin\theta  & \cos\theta \sin\varphi  & \cos\theta \cos\varphi  \end{array}
\right] 
\end{equation}

The direction of each spring and damper is denoted by $\boldsymbol{q_{ij}}   \in S^2$, where $S^2=\{\boldsymbol{q}\in \mathbb{R}^3|\ \left\|\boldsymbol{q}\right\|=1\}.$ The distance between the $i$th connection point (corresponding to the $i$th cable and $i$th quadrotor) and load mass center is denoted by $\boldsymbol{d_i}\in \mathbb{R}^3$. It is represented in the body-fixed frame of the load. In addition, spring length, spring constant and damping coefficient of the $j$th spring and damper of the $i$th cable are identified by $l_{ij}\in \mathbb{R}^+$, $K_{ij}$ and $b_{ij}$, respectively.

The $i$th quadrotor can provide a thrust force of $-f_i \boldsymbol{R_i} \boldsymbol{e_3}\in \mathbb{R}  ^{3}$, where  $f_i\in \mathbb{R}^+$ is the total thrust magnitude and $\boldsymbol{R_i}\in \mathbb{R}  ^{3\times 3\ }$ is the rotational matrix of the $i$th quadrotor. This thrust force is in the inertial frame and provides a torque $\boldsymbol{\tau_i}\in \mathbb{R}^{3}$ in its body frame. The thrust force $f_i$ and torque $\boldsymbol{\tau_i}$ are determined based on the rotor's angular velocities of the $i$th quadrotor. Finally, the effects of the disturbances and unmodeled dynamics on the equations of motion of the load position, the load angle, the $q_{ij}$, $l_{ij}$, and the rotation of the $i$th quadrotor, are demonstrated, respectively, by $\boldsymbol{\delta F_{r_l}}\in \mathbb{R}^{3}$, $\boldsymbol{\delta F_{\eta_l}}\in \mathbb{R}^{3}$, $\delta \boldsymbol{F_{q_{ij}}}\in \mathbb{R}^3$, $\delta F_{l_{ij}}\in \mathbb{R}$, and $\boldsymbol{\delta T_i}\in \mathbb{R}^{3}$. 

In general, the entire system can be seen as a single entity, with the center of mass of the load considered as the center of the system to which the body frame is attached. Therefore, the interactions between the flexible rope and the quadrotors, as well as between the quadrotors themselves, are internal interactions.   

\section{ Dynamical Modeling}
First, the positions of the $i$th quadrotor and $j$th mass of the $i$th cable (connecting the $i$th quadrotor to the load) are respectively given by 
\begin{align}
\boldsymbol{r_i} & = \boldsymbol{r_l} + \boldsymbol{R_l}\boldsymbol{d_i}-\sum^{n_i}_{a=1}{l_{ia}\boldsymbol{q_{ia}}}  \label{GrindEQ__2} &
\boldsymbol{r_{ij}} & = \boldsymbol{r_l} + \boldsymbol{R_l}\boldsymbol{d_i}-\sum^{n_i}_{a=j+1}{l_{ia}\boldsymbol{q_{ia}}} 
\end{align}
The velocities of the $i$th quadrotor and the $j$th mass of the $i$th cable are respectively
\begin{align}
\boldsymbol{\dot{r}_i} &= \boldsymbol{\dot{r}_l} + \boldsymbol{\dot{R}_l}\boldsymbol{d_i}-\sum^{n_i}_{a=1}{\left({\dot{l}}_{ia}\boldsymbol{q_{ia}} + l_{ia}\boldsymbol{\dot{q}_{ia}}\right)} \label{GrindEQ__4} &
\boldsymbol{\dot{r}_{ij}} & = \boldsymbol{\dot{r}_{l}}+\boldsymbol{\dot{R}_{l}}\boldsymbol{d_i} - \sum^{n_i}_{a=j+1}{({\dot{l}}_{ia}\boldsymbol{q_{ia}}+l_{ia}\boldsymbol{\dot{q}_{ia}})} 
\end{align}

and the kinematics equations $\boldsymbol{\dot{R}_{l}}$ and $\boldsymbol{\dot{q}_{ij}}$ are defined as
\begin{align}
\boldsymbol{\dot{R}_{l}} & = \boldsymbol{R_l}\boldsymbol{\widehat{\Omega}_l}
&
\boldsymbol{\dot{q}_{ij}} &= \boldsymbol{\omega_{ij}}\times\boldsymbol{q_{ij}} = \boldsymbol{\widehat{\omega}_{ij}}\boldsymbol{q_{ij}} \label{GrindEQ__7}
\end{align}


The angular velocity of the $j$th mass of the $i$th cable is represented by $\boldsymbol{\omega_{ij}}\in \mathbb{R}  ^{3}$ where $\boldsymbol{q_{ij}}\cdot  \boldsymbol{\omega_{ij}}=0$. Furthermore, $\boldsymbol{\Omega_l}\in \mathbb{R}  ^{3\ }$ is the angular velocity of the load. The hat notation is used for vector product, i.e.,  $\widehat{(\cdot)}:\mathbb{R}^3 \to \text{SO}(3)$ is defined as $\boldsymbol{\widehat{x}}\boldsymbol{y} = \boldsymbol{x}\times\boldsymbol{y}$ for $\boldsymbol{x},\boldsymbol{y}\in\mathbb{R}  ^{3}$. In other words, the hat map converts the vector $\boldsymbol{x}={\left[x_1,x_2,x_3\right]}^T$ to the skew-symmetric matrix $\widehat{\boldsymbol{x}}$ with the following form
\begin{equation}\label{GrindEQ__8}
\widehat{\boldsymbol{x}}=\left[ \begin{array}{ccc}
0 & {-x}_3 & x_2 \\ 
x_3 & 0 & -x_1 \\ 
{-x}_2 & x_1 & 0 \end{array}
\right] 
\end{equation}

The kinetic $T$ and potential $V$ energies of the system are calculated by  
\begin{align}
T&=\frac{1}{2}m_l{\left\|\boldsymbol{\dot{r}_{l}}\right\|}^2+\sum^n_{i=1}{\frac{1}{2}m_i{\left\|{\boldsymbol{\dot{r}}}_{\boldsymbol{i}}\right\|}^2}+\sum^n_{i=1}{\sum^{n_i}_{j=1}{\frac{1}{2}m_{ij}{\left\|{\boldsymbol{\dot{r}}}_{\boldsymbol{ij}}\right\|}^2}} +\frac{1}{2}{\boldsymbol\Omega}^{T}_{\boldsymbol{l}}{\boldsymbol{J}}_{\boldsymbol{l}}{\boldsymbol\Omega}_{\boldsymbol{l}}+\sum^n_{i=1}{\frac{1}{2}{\boldsymbol\Omega_{i}}^{T}{\boldsymbol{J}}_{\boldsymbol{i}}{\boldsymbol\Omega}_{\boldsymbol{i}}}
\label{GrindEQ__9}\\
V&=-m_lg{\boldsymbol{r}}^T_{\boldsymbol{l}} \boldsymbol{e_3}-\sum^n_{i=1}{m_ig{\boldsymbol{r}}^T_{\boldsymbol{i}} \boldsymbol{e_3}}+\sum^n_{i=1}{\sum^{n_i}_{j=1}{m_{ij}g{\boldsymbol{r}}^T_{\boldsymbol{ij}} \boldsymbol{e_3}}}+\sum^n_{i=1}{\sum^{n_i}_{j=1}{\frac{1}{2}K_{ij}\frac{\mathrm{\Delta }l_{ij}\left(\left|\mathrm{\Delta }l_{ij}\right|+\mathrm{\Delta }l_{ij}\right)}{4}}} \label{GrindEQ__10}\end{align}

The last term of the potential energy comes from potential energies of springs where it is zero as $\mathrm{\Delta }l_{ij}<0$ (shows that the cable is loose), and it is equal to $\frac{1}{2}K_{ij}\mathrm{\Delta }l_{ij} $ during the stretch of cable ($\mathrm{\Delta }l_{ij}>0$). Here, $\mathrm{\Delta }l_{ij}=l_{ij}-L_{ij}\ $ is the length change of the $j$th spring and damper in the $i$th cable. 
Then, the equations of motion of the aerial load transportation system are driven by the Euler-Lagrange method (see Appendix. \ref{app:a}) as follows:
\begin{equation}\label{GrindEQ__26}\begin{split} 
&M_T{\boldsymbol{\ddot{r}}}_{\boldsymbol{l}}-\sum^n_{i=1}{\sum^{n_i}_{j=1}{M_{cij}\left(l_{ij}\left( - {\widehat{\boldsymbol{q}}}_{\boldsymbol{ij}}{\boldsymbol{\dot{\omega }}}_{\boldsymbol{ij}} - \boldsymbol{\parallel }\boldsymbol{\omega_{ij}}{\boldsymbol{\parallel }}^{2}\boldsymbol{q_{ij}}\right)+2{\dot{l}}_{ij}\boldsymbol{\dot{q}_{ij}}+{\ddot{l}}_{ij}\boldsymbol{q_{ij}}\right)}} \\
&\quad -M_Tg \boldsymbol{e_3}+\sum^n_{i=1}{M_{qi}\boldsymbol{R_l}\left({\widehat{\boldsymbol\Omega}}^{2}_{\boldsymbol{l}}\boldsymbol{d_i} - {\widehat{\boldsymbol{d}}}_{\boldsymbol{i}}{\boldsymbol{\dot\Omega}}_{\boldsymbol{l}}\right)}=-\sum^n_{i=1}{f_i\boldsymbol{R_i} \boldsymbol{e_3}}+\delta F_{r_l}
\end{split}\end{equation}

\begin{equation}\label{GrindEQ__27}\begin{split} 
&\left({\boldsymbol{J}}_{\boldsymbol{l}}-\sum^n_{i=1}{M_{qi}}{\widehat{\boldsymbol{d}}}^{2}_{\boldsymbol{i}}\right){\boldsymbol{\dot\Omega}}_{\boldsymbol{l}} -\sum^n_{i=1}{\sum^{n_i}_{j=1}{M_{cij}\boldsymbol{{\hat{d}}_i}\boldsymbol{R_l}^{T}\left(l_{ij}\left( - {\widehat{\boldsymbol{q}}}_{\boldsymbol{ij}}{\boldsymbol{\dot{\omega }}}_{\boldsymbol{ij}} - \boldsymbol{\parallel }\boldsymbol{\omega_{ij}}{\boldsymbol{\parallel }}^{2}\boldsymbol{q_{ij}}\right)+2{\dot{l}}_{ij}\boldsymbol{\dot{q}_{ij}}+{\ddot{l}}_{ij}\boldsymbol{q_{ij}}\right)}}\\
&\quad +\sum^n_{i=1}{M_{qi}}{\widehat{\boldsymbol{d}}}_{\boldsymbol{i}}{\boldsymbol{R}}^{T}_{\boldsymbol{l}}{\boldsymbol{\ddot{r}}}_{\boldsymbol{l}} + {\widehat{\boldsymbol\Omega}}_{\boldsymbol{l}}\left({\boldsymbol{J}}_{\boldsymbol{l}} - \sum^n_{i=1}{M_{qi}}{\widehat{\boldsymbol{d}}}^{2}_{\boldsymbol{i}}\right){\boldsymbol\Omega}_{\boldsymbol{l}}- \sum^n_{i=1}{M_{qi}}g{\widehat{\boldsymbol{d}}}_{\boldsymbol{i}}{\boldsymbol{R}}^{T}_{\boldsymbol{l}} \boldsymbol{e_3}=-\sum^n_{i=1}{{f_i{\widehat{\boldsymbol{d}}}_{\boldsymbol{i}}{\boldsymbol{R}}^{T}_{\boldsymbol{l}}\boldsymbol{R_i}\boldsymbol{e_3}} }+ \delta F_{\eta_l}
\end{split}\end{equation}

\begin{equation}\label{GrindEQ__28}\begin{split}  
&M_{cij}l^2_{ij}\left({\widehat{\boldsymbol{q}}}_{\boldsymbol{ij}}{\boldsymbol{\dot{\omega }}}_{\boldsymbol{ij}} + \boldsymbol{\parallel }\boldsymbol{\omega_{ij}}{\boldsymbol{\parallel }}^{2}\boldsymbol{q_{ij}}\right)-\sum^{n_i}_{k=1 \atop k\neq j}{M_{cij}l_{ij}l_{ik}{\widehat{\boldsymbol{q}}}^{2}_{\boldsymbol{ij}}\left({\widehat{\boldsymbol{q}}}_{\boldsymbol{ik}}{\boldsymbol{\dot{\omega }}}_{\boldsymbol{ik}} + \parallel {\boldsymbol{\omega }}_{\boldsymbol{ik}}{\boldsymbol{\parallel }}^{2}{\boldsymbol{q}}_{\boldsymbol{ik}}\right)} -M_{cij}l^2_{ij}\boldsymbol{\parallel }\boldsymbol{\dot{q}_{ij}}{\boldsymbol{\parallel }}^{2}\boldsymbol{q_{ij}}\\
&\quad \quad +2\sum^{n_i}_{k=1}{M_{cij}l_{ij}{\dot{l}}_{ik}{\widehat{\boldsymbol{q}}}^{2}_{\boldsymbol{ij}}{\boldsymbol{\dot{q}}}_{\boldsymbol{ik}}}+\sum^{n_i}_{k=1 \atop k\neq j }{M_{cij}l_{ij}{\ddot{l}}_{ik}{\widehat{\boldsymbol{q}}}^{2}_{\boldsymbol{ij}}{\boldsymbol{q}}_{\boldsymbol{ik}}} -M_{cij}{\dot{l}}^2_{ij}{\widehat{\boldsymbol{q}}}^{2}_{\boldsymbol{ij}}\boldsymbol{q_{ij}}\\
&\quad \quad -M_{cij}l_{ij}{\widehat{\boldsymbol{q}}}^{2}_{\boldsymbol{ij}}\left({\boldsymbol{\ddot{r}}}_{\boldsymbol{l}} + \boldsymbol{R_l}\left({\widehat{\boldsymbol\Omega}}^{2}_{\boldsymbol{l}}\boldsymbol{d_i} - {\widehat{\boldsymbol{d}}}_{\boldsymbol{i}}{\boldsymbol{\dot\Omega}}_{\boldsymbol{l}}\right)\right)+M_{cij}gl_{ij}{\widehat{\boldsymbol{q}}}^{2}_{\boldsymbol{ij}} \boldsymbol{e_3}=l_{ij}{f_i{\widehat{\boldsymbol{q}}}^{2}_{\boldsymbol{ij}}\boldsymbol{R_i}\boldsymbol{e_3}}+\delta F_{q_{ij}} 
\end{split}\end{equation}

\begin{equation}\label{GrindEQ__29}\begin{split} 
&\sum^{n_i}_{k=1}{M_{cij}{\ddot{l}}_{ik}{\boldsymbol{q}}^{T}_{\boldsymbol{ij}}{\boldsymbol{q}}_{\boldsymbol{ik}}}-\sum^{n_i}_{k=1 \atop k\neq j}{M_{cij}{\boldsymbol{q}}^{T}_{\boldsymbol{ij}}\left(2{\dot{l}}_{ik}{\boldsymbol{q}}_{\boldsymbol{ik}} + l_{ik}\left({\widehat{\boldsymbol{q}}}_{\boldsymbol{ik}}{\boldsymbol{\dot{\omega }}}_{\boldsymbol{ik}} + \boldsymbol{\parallel }{\boldsymbol{\omega }}_{\boldsymbol{ik}}{\boldsymbol{\parallel }}^{2}{\boldsymbol{q}}_{\boldsymbol{ik}}\right)\right)} \\
&\quad +M_{cij}g{\boldsymbol{q}}^{T}_{\boldsymbol{ij}} \boldsymbol{e_3}-M_{cij}{\boldsymbol{q}}^{T}_{\boldsymbol{ij}}\left({\boldsymbol{\ddot{r}}}_{\boldsymbol{l}} + \boldsymbol{R_l}\left({\widehat{\boldsymbol\Omega}}^{2}_{\boldsymbol{l}}\boldsymbol{d_i} - {\widehat{\boldsymbol{d}}}_{\boldsymbol{i}}{\boldsymbol{\dot\Omega}}_{\boldsymbol{l}}\right)\right) -M_{cij}l_{ij}{\boldsymbol{\dot{q}}}^{T}_{\boldsymbol{ij}}\boldsymbol{\dot{q}_{ij}}\\
&\quad +K_{ij}\left(l_{ij}-L_{ij}\right)\frac{1+\sign\left(l_{ij}-L_{ij}\right)}{2} +b_{ij}{\dot{l}}_{ij}\frac{1+\sign\left(l_{ij}-L_{ij}\right)}{2}=-{f_i{\boldsymbol{q}}^{T}_{\boldsymbol{ij}}\boldsymbol{R_i}\boldsymbol{e_3}}+\delta F_{l_{ij}}
\end{split}\end{equation}

\begin{equation}\label{GrindEQ__22}
{\boldsymbol{J}}_{\boldsymbol{i}}{\boldsymbol{\dot\Omega}}_{\boldsymbol{i}} - {\widehat{\boldsymbol\Omega}}_{\boldsymbol{i}}{\boldsymbol{J}}_{\boldsymbol{i}}{\boldsymbol\Omega}_{\boldsymbol{i}} = {\boldsymbol{\tau}}_{\boldsymbol{i}}+\delta T_i 
\end{equation}

Since the rotation part of quadrotors (Eq. \ref{GrindEQ__22}) is decoupled from the other part of the dynamic, equations \eqref{GrindEQ__26} to \eqref{GrindEQ__29} in the matrix form will be
\begin{equation}\label{GrindEQ__30}
\boldsymbol{M}\boldsymbol{\dot{X}}+\boldsymbol{C}=\boldsymbol{P}+\boldsymbol{P_\delta} 
\end{equation}
where $\boldsymbol{X}=\begin{bmatrix}
\boldsymbol{\dot{r_l}} & \boldsymbol{\Omega_l} & {\boldsymbol{\omega }}_{\boldsymbol{1}\boldsymbol{j}} & \cdots  & {\boldsymbol{\omega }}_{\boldsymbol{nj}} & \dot{l}_{1j} & \cdots  & \dot{l}_{nj}
\end{bmatrix}^T  \in\mathbb{R}^d$, $\boldsymbol{M}  \in\mathbb{R}^{d\times d}$, $\boldsymbol{\mathrm{C}}  \in\mathbb{R}^d$, $\boldsymbol{P_\delta}\in\mathbb{R}^d$, and $\boldsymbol{P}  \in\mathbb{R}^{d}$ with $d=6+4nm$ are the state vector, mass matrix, Coriolis matrix, disturbance vector, and control input, respectively. These matrices are presented in detail in the Appendix. \ref{app:b}.

\section{ Controller Design}

\begin{figure} \centering \includegraphics*[width=5.0in]{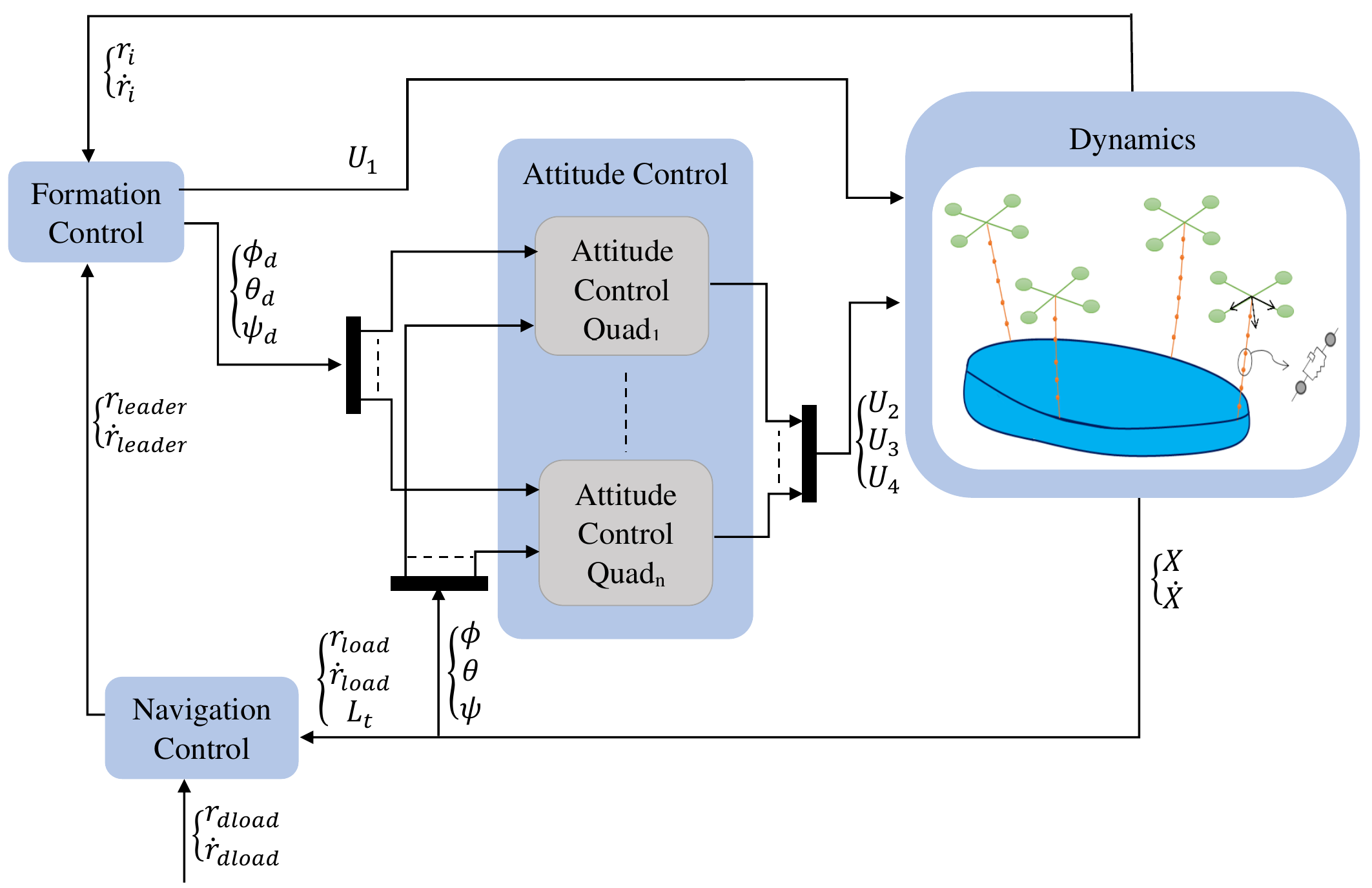}
	\caption{Block diagram of the proposed control system for multi-agent transportation of a rigid body.}
	\label{fig2}
\end{figure}

In this section, a control framework is proposed for multi-agent aerial transportation of a rigid body to the desired trajectory. This control framework is able to eliminate problems such as payload fluctuations, roping cables and collision of the quadrotors. The control strategy consists of three parts: the first part is the navigation control which receives feedback directly from the states of load and aids to navigate the system to the desired position with proper velocity. The next part is the formation control which guarantees that agents (quadrotors) get in proper formation with respect to their neighbors and the leader. This can reduce the swing of the load and prevent the collision of quadrotors. The desired angles and thrust force of each quadrotor are provided from the formation control part. Finally, attitude control as the third part generates agents control signal to stabilize the attitude of each quadrotor based on the desired angles. The attitude control part has internal feedback to stabilize each quadrotor. Figure.~\ref{fig2} shows the block diagram of the proposed transportation control system.

\subsection{ Navigation Control}

The navigation control guides the system to the desired position and velocity by creating a proper center of the formation. The center of formation (leader) is an imaginary point or is the center of the payload which must guarantee that the load gets the desired trajectory and also the group move with a proper velocity. Let's define the position and velocity of the leader as  
\begin{equation}\label{GrindEQ__59}\begin{aligned}  {\boldsymbol{r}}_{\boldsymbol{leader}} = {\boldsymbol{r}}_{\boldsymbol{load}} + {\boldsymbol{k}}_{\boldsymbol{r}}\left({\boldsymbol{r}}_{\boldsymbol{dload}} - {\boldsymbol{r}}_{\boldsymbol{load}}\right) + {\boldsymbol{k}}_{\boldsymbol{i}}\int{\left({\boldsymbol{r}}_{\boldsymbol{dload}} - {\boldsymbol{r}}_{\boldsymbol{load}}\right)\boldsymbol{dt}} + {\left[0\ 0\ L_t\right]}^T
\end{aligned}\end{equation}
\begin{equation}\label{GrindEQ__60}\begin{aligned} 
{\boldsymbol{v}}_{\boldsymbol{leader}} = {\boldsymbol{v}}_{\boldsymbol{load}} + {\boldsymbol{k}}_{\boldsymbol{v}}\left({\boldsymbol{v}}_{\boldsymbol{dload}} - {\boldsymbol{v}}_{\boldsymbol{load}}\right) 
\end{aligned}\end{equation}
where ${\boldsymbol{r}}_{\boldsymbol{load}}$\textbf{, }${\boldsymbol{r}}_{\boldsymbol{dload}}$, ${\boldsymbol{v}}_{\boldsymbol{load}}$ and ${\boldsymbol{v}}_{\boldsymbol{dload}}$ express the position, desired position, velocity and desired velocity of the load, respectively. Furthermore, the last term in \eqref{GrindEQ__59} includes the length of cables $L_t$ to make the leader elevation equal to that of followers. Diagonal positive matrices of ${\boldsymbol{k}}_{\boldsymbol{r}}$, ${\boldsymbol{k}}_{\boldsymbol{i}}$ and ${\boldsymbol{k}}_{\boldsymbol{v}}$ are tuned to achieve proper performance of the system.

\subsection{ Formation Control}

Formation control is designed to avoid collision of quadrotors and twisting cables together, and also distribute the rigid body weight between all quadrotors uniformly.

\subsubsection{ Preliminary }

First, a brief explanation of multi-agent systems was given as well as a preliminary of such systems. In general, multi-agent systems have broad applications such as UAVs \cite{Rezaei1}, and they can be applied along with a combination of different methods like neural networks and optimization approaches (\cite{pedram}, \cite{saleh}). Consider a group of $n$ followers and a leader, in which the leader tracks a desired trajectory and followers move along with the leader. The communication of followers with each other and the leader is exposed by a graph $G=(\upsilon ,\varepsilon )$ where $\upsilon =\{{\upsilon }_1\ ,{\upsilon }_2\ ,...,{\upsilon }_n\}$ and $\varepsilon$ denote node and edge sets, respectively. In the graph, each edge $({\upsilon }_j,{\upsilon }_i)\in \ \varepsilon $ with a specified weight $a_{ij}\ \ge \ 0$ is an ordered pair of connected nodes. The neighbor set of the agent $i$ is represented as ${\boldsymbol{N}}_{\boldsymbol{i}}=\{\ {\upsilon }_j\ \ :\ ({\upsilon }_j,{\upsilon }_i)\in \ \varepsilon \}$. It means that the agent $i$ and $j$ are neighbors and the agent $i$ sends the information to agent $j$, if there is an edge between them. Furthermore, in the spanning tree, $a_{ij}=\ a_{ji}$, and no agent transfers the information to itself i.e.$\ a_{ii}=0$.

The adjacency matrix is given by 
\begin{equation}\label{GrindEQ__57}\begin{aligned} 
\boldsymbol{A}\mathrm{=}\left[ \begin{array}{cccc}
0 & a_{12} & \cdots  & a_{1n} \\ 
a_{21} & 0 & \cdots  & a_{2n} \\ 
\vdots  & \vdots  & \ddots  & \vdots  \\ 
a_{n1} & a_{n2} & \cdots  & 0 \end{array}
\right] 
\end{aligned}\end{equation}

The degree matrix $D$ $  \in \mathbb{R}  ^{n\times n}$ is defined as a diagonal matrix where $d_i=\sum^n_{j=1}{a_{ij}}$ ($d _{i}$ is the in-degree of ${\upsilon }_i$). Based on the adjacency and degree matrices, the Laplacian matrix is determined by $\boldsymbol{L} = \boldsymbol{D}-\boldsymbol{A}\in \mathbb{R}  ^{n\times n}$ where
\begin{equation}\label{GrindEQ__58}\begin{aligned} 
l_{\mathrm{ij}}=\left\{ \begin{array}{c}
d_i\ \ \ \ \ \ \ \ \ \ \ \ \ \ \ \ \ \ \ \ \ \ \ \ i=j \\ 
-1 \ \ \ \  \ \ \left(i,j\right)\in E,\ i\neq j \\ 
0\ \ \ \ \ \ \ \ \ \ \ \ \ \ \ \ otherwise \end{array}
\right. 
\end{aligned}\end{equation}

In addition, the relationship between the leader and followers is represented $by\ \boldsymbol{B}=diag\{b_1\ ,b_2\ ,...,b_n\}.$

\subsubsection{Formation Control Design}

To denote the dynamics of each agent in $x$, $y$ and $z$ coordinates, first for a quadrotor define
\begin{equation}\label{GrindEQ__61}\begin{aligned}
f\left(X,U\right)=\left[ \begin{array}{c}
\dot{x} \\ 
-\frac{U_1}{m}(sin\psi \sin\phi +cos\phi \cos\psi \sin\theta )+{\delta }_x \\ 
\dot{y} \\ 
\frac{U_1}{m}(sin\phi \cos\psi -cos\phi \sin\psi \sin\theta )+{\delta }_y \\ 
\dot{z} \\ 
g-\frac{U_1}{m}cos\phi cos\theta +{\delta }_z \end{array}
\right] 
\end{aligned}\end{equation}
where $\dot{x}$, $\dot{y}$ and $\dot{z}$ are its velocities in the inertia coordinates; $\phi $, $\theta $ and $\psi $ are its Euler angles in the inertia coordinates; $m$ is its mass; $g$ is the gravitational acceleration; $U_1$ is its thrust; ${\delta }_x$, ${\delta }_y$ and ${\delta }_z$ represent the effect of uncertainties and disturbances due to unmodeled dynamics, unknown payload, and external disturbances.

The Euler angles of $\phi $, $\theta $ and $\psi $ are respectively substituted by ${\phi }_d$, ${\theta }_d$ and ${\psi }_d$ which ${\psi }_d$ is always supposed zero, thus
\begin{equation}\label{GrindEQ__62}\begin{aligned}
f\left(X,U\right)=\left[ \begin{array}{c}
\dot{x} \\ 
-\frac{U_1}{m}cos{\phi }_d\sin{\theta }_d+{\delta }_x \\ 
\dot{y} \\ 
\frac{U_1}{m}sin{\phi }_d\ +{\delta }_y \\ 
\dot{z} \\ 
g-\frac{U_1}{m}cos{\phi }_d\cos{\theta }_d+{\delta }_z \end{array}
\right] 
\end{aligned}\end{equation}

The recent equation can be rewritten as

\begin{equation}\label{GrindEQ__63}\begin{aligned}
f\left(X,U\right)={\left[
\dot{x}=v_x \quad 
\ddot{x}=u_x+{\delta }_x \quad
\dot{y}=v_y \quad
\ddot{y}=u_y+{\delta }_y \quad
\dot{z}=v_z \quad
\ddot{z}=u_z+{\delta }_z
\right]}^T 
\end{aligned}\end{equation}

Hence, formation control is designed for a system that each of its agents has 3D double integrator dynamics. Now, define

\begin{align}\label{GrindEQ__64}
\boldsymbol{r_i}&={\left[ 
x_i \quad y_i \quad z_i 
\right]}^T &
\boldsymbol{v_i}={\left[
v_{xi} \quad v_{yi} \quad v_{zi}
\right]}^T & \quad
\boldsymbol{u_i}={\left[
u_{xi} \quad u_{yi} \quad u_{zi}
\right]}^T &
{\boldsymbol{\mathit{\Delta}}}_{\boldsymbol{i}}={\left[
{\delta }_{xi} \quad {\delta }_{yi} \quad {\delta }_{zi}
\right]}^T 
\end{align}
where $\boldsymbol{r_i}, \boldsymbol{v_i}$, $\boldsymbol{u_i}$ and ${\boldsymbol{\mathit{\Delta}}}_{\boldsymbol{i}}\boldsymbol{\ }$are vectors of the position, velocity, control input and disturbances of the $i$th agent, respectively. Thus, the motion equation of the $i$th agent is given as 
\begin{equation}\label{GrindEQ__65}
\left\{ \begin{array}{l}
\boldsymbol{\dot{r}_i} = \boldsymbol{v_i} \\ 
\boldsymbol{\dot{v}_i} = \boldsymbol{u_i} + {\boldsymbol{\mathit{\Delta}}}_{\boldsymbol{i}} \end{array}
\right. 
\end{equation}

The formation task is a leader-follower based problem and the leader dynamics is given by
\begin{equation}\label{GrindEQ__66}
\left\{ \begin{array}{l}
\boldsymbol{\dot{r}_{l}} = \boldsymbol{v_l} \\ 
\boldsymbol{\dot{v}_l} = {\boldsymbol{U}}_{\boldsymbol{l}} \end{array}
\right. 
\end{equation}

\textbf{Assumption 1}. The graph has at least one directed spanning tree.

The multi-agent system requires to reach the desired formation, therefore, the error position and velocity of the $i$th agent is defined as

\begin{align}
\boldsymbol{e_{ri}} &=\sum^n_{j=1}{a_{ij}\left(\boldsymbol{r_i} - \boldsymbol{r_{j}} - \boldsymbol{H_{ij}}\right)+b_i(\boldsymbol{r_i} - \boldsymbol{r_l} - \boldsymbol{H_{il}})} \label{GrindEQ__67} &
\boldsymbol{e_{vi}}=\sum^n_{j=1}{a_{ij}\left(\boldsymbol{v_i} - \boldsymbol{v_{j}}\right)+b_i(\boldsymbol{v_i} - \boldsymbol{v_l})} 
\end{align}
where $\boldsymbol{\ }{\boldsymbol{e}}_{\boldsymbol{ri}}={\left[ \begin{array}{ccc}
e_{xi} & e_{yi} & e_{zi} \end{array}
\right]}^T$, $\boldsymbol{\ }{\boldsymbol{e}}_{\boldsymbol{vi}}={\left[ \begin{array}{ccc}
e_{xi} & e_{yi} & e_{zi} \end{array}
\right]}^T$, $\boldsymbol{H_{ij}}={\left[ \begin{array}{ccc}
h_{xij} & h_{yij} & h_{zij}\end{array}
\right]}^T$ and $\boldsymbol{H_{il}}={\left[ \begin{array}{ccc}
h_{xil} & h_{yil} \end{array}
\right.}$
${\left. \begin{array}{ccc} h_{zil}  \end{array}
\right]}^T$ express the position error, velocity error, desired distance of the $i$th agent from the $j$th agent and the desired distance of the $i$th agent from the leader, respectively. Based on \textbf{Assumption 1}, the error functions for the general multi-agent system are obtained as 
\begin{equation}\label{GrindEQ__69}\begin{aligned}
\boldsymbol{E} = \left(\left(\mathcal{L} + \boldsymbol{B}\right)\boldsymbol{\otimes }\boldsymbol{I}_{\boldsymbol{3}}\right)\left(\boldsymbol{r} - \boldsymbol{H}\right) - \left(\boldsymbol{B}\boldsymbol{\otimes }{\boldsymbol{I}}_{\boldsymbol{3}}\right)\left({\boldsymbol{1}}_{\boldsymbol{n}}\boldsymbol{\otimes }\boldsymbol{r_l} - {\boldsymbol{H}}_{\boldsymbol{l}}\right)
\end{aligned}\end{equation}
\begin{equation}\label{GrindEQ__70}\begin{aligned} 
{\boldsymbol{E}}_{\boldsymbol{v}} = \left(\left(\mathcal{L} + \boldsymbol{B}\right)\boldsymbol{\otimes }\boldsymbol{I}_{\boldsymbol{3}}\right)\boldsymbol{v} - \left(\boldsymbol{B}\boldsymbol{\otimes }{\boldsymbol{I}}_{\boldsymbol{3}}\right)\boldsymbol{v_l} 
\end{aligned}\end{equation}
where$\boldsymbol{\ }\boldsymbol{E} = {\left[ \begin{array}{cccc}
 \boldsymbol{e_1} &  \boldsymbol{e_2} & \boldsymbol{\cdots } & {\boldsymbol{e}}_{\boldsymbol{n}} \end{array}
\right]}^T  \in\mathbb{R}^{\mathrm{3n}}$, $\boldsymbol{\ }{\boldsymbol{E}}_{\boldsymbol{v}} = {\left[ \begin{array}{cccc}
{\boldsymbol{e}}_{\boldsymbol{v}\boldsymbol{1}} & {\boldsymbol{e}}_{\boldsymbol{v}\boldsymbol{2}} & \boldsymbol{\cdots } & {\boldsymbol{e}}_{\boldsymbol{vn}} \end{array}
\right]}^T  \in\mathbb{R}^{\mathrm{3n}}$, $\mathcal{L}$ is Laplacian matrix, $\boldsymbol{\otimes }$ represents the Kronecker product, $\boldsymbol{B} = {\left[ \begin{array}{cccc}
 b_1 &  b_2 \end{array}
\right.}$
${\left. \begin{array}{ccc} &\cdots & b_n \end{array}
\right]}^T\in \mathbb{R}^n$, ${\boldsymbol{r}}_{\boldsymbol{f}} = {\left[ \begin{array}{cccc}
{\boldsymbol{r}}_{\boldsymbol{1}} & {\boldsymbol{r}}_{\boldsymbol{2}} & \boldsymbol{\cdots } & {\boldsymbol{r}}_{\boldsymbol{n}} \end{array}
\right]}^T\in \mathbb{R}^{3n}$, $\boldsymbol{\ }{\boldsymbol{v}}_{\boldsymbol{f}} = {\left[ \begin{array}{cccc}
{\boldsymbol{v}}_{\boldsymbol{1}} & {\boldsymbol{v}}_{\boldsymbol{2}} \end{array}
\right.}$
${\left. \begin{array}{ccc} &\boldsymbol{\cdots } & {\boldsymbol{v}}_{\boldsymbol{n}} \end{array}
\right]}^T\in \mathbb{R}^{3n}, \boldsymbol{H} = {\left[ \begin{array}{cccc}
{\boldsymbol{H}}_{\boldsymbol{1}\boldsymbol{j}} & {\boldsymbol{H}}_{\boldsymbol{2}\boldsymbol{j}} & \boldsymbol{\cdots } & {\boldsymbol{H}}_{\boldsymbol{nj}} \end{array}
\right]}^T\in \mathbb{R}^{3n}$ and $H_l={\left[ \begin{array}{cccc}
H_{1l} & H_{2l} & \cdots & H_{nl} \end{array}
\right]}^T\in \mathbb{R}^{3n}$.

By differentiating \eqref{GrindEQ__70}, we have
\begin{equation}\label{GrindEQ__71}
{\boldsymbol{\dot{E}}}_{\boldsymbol{v}} = \left(\left(\mathcal{L} + \boldsymbol{B}\right)\boldsymbol{\otimes }\boldsymbol{I}_{\boldsymbol{3}}\right)\boldsymbol{U} + \left(\left(\mathcal{L} + \boldsymbol{B}\right)\boldsymbol{\otimes }\boldsymbol{I}_{\boldsymbol{3}}\right)\boldsymbol{\mathit{\Delta}} - \left(\boldsymbol{B}\boldsymbol{\otimes }{\boldsymbol{I}}_{\boldsymbol{3}}\right){\boldsymbol{u}}_{\boldsymbol{l}} 
\end{equation}
with
\begin{equation}
\left\{ \begin{array}{c}
\boldsymbol{U} = {\left[ \begin{array}{cccc}
{\boldsymbol{U}}_{\boldsymbol{1}} & {\boldsymbol{U}}_{\boldsymbol{2}} & \boldsymbol{\cdots } & {\boldsymbol{U}}_{\boldsymbol{n}} \end{array}
\right]}^T\in \mathbb{R}^{3n} \\ 
\boldsymbol{\mathrm{\Delta }} = {\left[ \begin{array}{cccc}
{\boldsymbol{\mathrm{\Delta }}}_{\boldsymbol{1}} & {\boldsymbol{\mathrm{\Delta }}}_{\boldsymbol{2}} & \boldsymbol{\cdots } & {\boldsymbol{\mathrm{\Delta }}}_{\boldsymbol{n}} \end{array}
\right]}^T\in \mathbb{R}^{3n} \end{array}
\right.
\end{equation} 

To achieve the desired formation, we employ the sliding mode approach and define the sliding surface as
\begin{equation}\label{GrindEQ__72}\begin{aligned} 
\boldsymbol{S} = \boldsymbol{\mathit{\Lambda}}\boldsymbol{E} + \boldsymbol{E}_{\boldsymbol{v}}
\end{aligned}\end{equation}
where $\boldsymbol{\Lambda}=diag({\lambda }_{x1},{\lambda }_{y1},{\lambda }_{z1},\cdots ,{\lambda }_{xn},{\lambda }_{yn},{\lambda }_{zn})\in \mathbb{R}^{3n\times 3n}$ with ${\lambda }_{xi}, {\lambda }_{yi}, {\lambda }_{zi}>0$ and $\boldsymbol{\mathrm{S}}\mathrm{=}{\left[ \begin{array}{cccc}
{\boldsymbol{S}}_{\boldsymbol{1}} & {\boldsymbol{S}}_{\boldsymbol{2}} & \boldsymbol{\cdots } & {\boldsymbol{S}}_{\boldsymbol{n}} \end{array}
\right]}^T\in \mathbb{R}^{3n}$ with ${\boldsymbol{S}}_{\boldsymbol{i}}={\left[ \begin{array}{ccc}
s_{xi} & s_{yi} & s_{zi} \end{array}
\right]}^T\in \mathbb{R}^3.$ Differentiating \eqref{GrindEQ__72} is defined as
\begin{equation}\label{GrindEQ__73}\begin{aligned} 
\boldsymbol{\dot{S}} = &\boldsymbol{\mathit{\Lambda}}{\boldsymbol{E}}_{\boldsymbol{v}} + \boldsymbol{\dot{E}}_{\boldsymbol{v}} = \boldsymbol{\mathit{\Lambda}}\left\{\left(\left(\mathcal{L} + \boldsymbol{B}\right)\boldsymbol{\otimes }\boldsymbol{I}_{\boldsymbol{3}}\right){\boldsymbol{v}}_{\boldsymbol{f}} - \left(\boldsymbol{B}\boldsymbol{\otimes }{\boldsymbol{I}}_{\boldsymbol{3}}\right)\boldsymbol{v_l}\right\} + \left(\left(\mathcal{L} + \boldsymbol{B}\right)\boldsymbol{\otimes }\boldsymbol{I}_{\boldsymbol{3}}\right)\boldsymbol{U}&\textbf{\textit{\newline }}\\& + \left(\left(\mathcal{L} + \boldsymbol{B}\right)\boldsymbol{\otimes }\boldsymbol{I}_{\boldsymbol{3}}\right)\boldsymbol{\mathit{\Delta}} - \left(\boldsymbol{B}\boldsymbol{\otimes }{\boldsymbol{I}}_{\boldsymbol{3}}\right){\boldsymbol{u}}_{\boldsymbol{l}} 
\end{aligned}\end{equation}

Finally, the control input is given by
\begin{equation}\label{GrindEQ__74}\begin{aligned} 
\boldsymbol{U} = &{ - \left(\left(\mathcal{L} + \boldsymbol{B}\right)\boldsymbol{\otimes }\boldsymbol{I}_{\boldsymbol{3}}\right)}^{ - \boldsymbol{1}}\left\{\boldsymbol{\mathit{\Lambda}}\left\{\left(\left(\mathcal{L} + \boldsymbol{B}\right)\boldsymbol{\otimes }\boldsymbol{I}_{\boldsymbol{3}}\right){\boldsymbol{v}}_{\boldsymbol{f}} - \left(\boldsymbol{B}\boldsymbol{\otimes }{\boldsymbol{I}}_{\boldsymbol{3}}\right)\boldsymbol{v_l}\right\} - \left(\boldsymbol{B}\boldsymbol{\otimes }{\boldsymbol{I}}_{\boldsymbol{3}}\right){\boldsymbol{u}}_{\boldsymbol{l}}\right.\\&\left. + \left\{\left(\left(\mathcal{L} + \boldsymbol{B}\right)\boldsymbol{\otimes }\boldsymbol{I}_{\boldsymbol{3}}\right)\boldsymbol{K} + \boldsymbol{M}\right\}\boldsymbol{sgn}\boldsymbol{(}\boldsymbol{S}\boldsymbol{)}\right\} 
\end{aligned}\end{equation}
where $\boldsymbol{M}=diag\left({\mu }_{x1},{\mu }_{y1},{\mu }_{z1},\cdots ,{\mu }_{xn},{\mu }_{yn},{\mu }_{zn}\right)  \in\mathbb{R}^{\mathrm{3n\times 3n}}$, $\boldsymbol{K}={\left[ \begin{array}{ccc} k_{x1} & k_{y1} & k_{z1} \end{array} \right.}$
${\left. \begin{array}{cccc} \cdots & k_{xn} & k_{yn} & k_{zn} \end{array} \right]}^T  \in\mathbb{R}^{\mathrm{3n\times 3n}}$ is the upper bound of $\boldsymbol{\mathit{\Delta}}$ and$\boldsymbol{\ }\boldsymbol{sgn}\left(\boldsymbol{S}\right) = {\left[ \begin{array}{c} \boldsymbol{sgn}\left({\boldsymbol{S}}^{T}_{\boldsymbol{1}}\right) \end{array} \right.}$
${\left. \begin{array}{ccc} \boldsymbol{sgn}\left({\boldsymbol{S}}^{T}_{\boldsymbol{2}}\right) & \boldsymbol{\cdots } & \boldsymbol{sgn}\left({\boldsymbol{S}}^{T}_{\boldsymbol{n}}\right) \end{array}
\right]}^{T}  \in\mathbb{R}^{\mathrm{3n}}$.

\textbf{\textit{Theorem} \textit{1}}. The multi-agent system with the followers dynamics \eqref{GrindEQ__65} and leader dynamics \eqref{GrindEQ__66}, control input \eqref{GrindEQ__74}, sliding surface \eqref{GrindEQ__72} and a directed spanning tree is stable, if each components of the matrices of $\boldsymbol{M}$ and $\boldsymbol{K}$ satisfies ${\mu }_{xi},\ {\mu }_{yi},{\mu }_{zi},k_{xi},\ k_{yi},k_{zi}>0$. 

\textbf{\textit{Proof}: }Consider $V'=\frac{1}{2}{\boldsymbol{S}}^{T}\boldsymbol{S}=\frac{1}{2}{\mathit{\Sigma}}^N_{i=1}\boldsymbol{S_i}^2$ as the Lyapunov function candidate. Then, its time differentiation will be 
\begin{equation}\label{GrindEQ__75}\begin{aligned} 
\dot{V}' = &{\boldsymbol{S}}^{T}\boldsymbol{\dot{S}} = {\boldsymbol{S}}^{T}\left\{\boldsymbol{\mathit{\Lambda}}\left\{\left(\left(\mathcal{L} + \boldsymbol{B}\right)\boldsymbol{\otimes }\boldsymbol{I}_{\boldsymbol{3}}\right){\boldsymbol{v}}_{\boldsymbol{f}} - \left(\boldsymbol{B}\boldsymbol{\otimes }{\boldsymbol{I}}_{\boldsymbol{3}}\right)\boldsymbol{v_l}\right\} + \left(\left(\mathcal{L} + \boldsymbol{B}\right)\boldsymbol{\otimes }\boldsymbol{I}_{\boldsymbol{3}}\right)\boldsymbol{U}\right.&\textbf{\textit{\newline }}\\&\left. + \left(\left(\mathcal{L} + \boldsymbol{B}\right)\boldsymbol{\otimes }\boldsymbol{I}_{\boldsymbol{3}}\right)\boldsymbol{\mathit{\Delta}} - \left(\boldsymbol{B}\boldsymbol{\otimes }{\boldsymbol{I}}_{\boldsymbol{3}}\right){\boldsymbol{u}}_{\boldsymbol{l}}\right\} 
\end{aligned}\end{equation}

By substituting \eqref{GrindEQ__73} into \eqref{GrindEQ__75} $\dot{V}'$, one can write 
\begin{equation}\label{GrindEQ__76}\begin{aligned} 
\dot{V}' = {\boldsymbol{S}}^{T}\left\{\left(\left(\mathcal{L} + \boldsymbol{B}\right)\boldsymbol{\otimes }\boldsymbol{I}_{\boldsymbol{3}}\right)\boldsymbol{\mathit{\Delta}} - \left\{\left(\left(\mathcal{L} + \boldsymbol{B}\right)\boldsymbol{\otimes }\boldsymbol{I}_{\boldsymbol{3}}\right)\boldsymbol{K} + \boldsymbol{M}\right\}\boldsymbol{sgn}\boldsymbol{(}\boldsymbol{S}\boldsymbol{)}\right\}
\end{aligned}\end{equation}
where $\boldsymbol{K}$ is the upper bound of $\boldsymbol{\mathit{\Delta}}$. Define $\boldsymbol{Q} = \left(\mathcal{L} + \boldsymbol{B}\right)\boldsymbol{\otimes }\boldsymbol{I}_{\boldsymbol{3}}$, $\dot{V}'$ can be rewritten as
\begin{equation}\label{GrindEQ__77}\begin{aligned} 
\dot{V}'\boldsymbol{\le }{\boldsymbol{S}}^{T}\left\{\boldsymbol{QK} - \left\{\boldsymbol{QK} + \boldsymbol{M}\right\}\boldsymbol{sgn}\left(\boldsymbol{S}\right)\right\}=\sum^N_{i=1}{\left(\boldsymbol{Q_iK_i({S_i}-|S_i|)-M_i|S_i|}\right)} 
\end{aligned}\end{equation}

Hence, choosing ${\mu }_{xi}, {\mu }_{yi}, {\mu }_{zi},k_{xi}, k_{yi}, k_{zi}>0$ such that $\dot{V}'\boldsymbol{<}\boldsymbol{0}$ and as regards $V'> 0$ thus the closed-loop control system is asymptotically stable.       $ \hfill\blacksquare $

\subsection{ Attitude Control}

The aim of attitude control is the stabilization of each quadrotor. Here, a Proportional-Integrator-Derivative controller is used for stabilization of each quadrotor \cite{15npp} and \cite{24pp} 
\begin{equation}\label{GrindEQ__78}\begin{aligned} 
U_2=K_{p\phi }\left({\phi }_d-\phi \right)-K_{d\phi }\dot{\phi }+K_{i\phi }\int \left({\phi }_d-\phi \right)
\end{aligned}\end{equation}
\begin{equation}\label{GrindEQ__79}\begin{aligned}  
U_3=K_{p\theta }\left({\theta }_d-\theta \right)-K_{d\theta }\dot{\theta }+K_{i\theta }\int \left({\theta }_d-\theta \right)
\end{aligned}\end{equation}
\begin{equation}\label{GrindEQ__80}\begin{aligned} 
U_4=K_{p\psi }\left({\psi }_d-\psi \right)-K_{d\psi }\dot{\psi }+K_{i\psi }\int ({\psi }_d-\psi ) 
\end{aligned}\end{equation}
where $K _{p\varphi }$, $K _{p\theta }$ and $K _{p\psi }$ are the proportional coefficients for control of the angles $\varphi $, $\theta$ and $\psi $, respectively; $K _{d\varphi }$, $K _{d\theta }$ and $K _{d\psi }$ are the derivative coefficients for control of the angular velocities; the integrator coefficients are indicated by $K _{i\varphi }$, $K _{i\theta }$ and $K _{i\psi }$; ${\varphi }_d$, $\theta  _{d}$ and ${\psi }_d$ are the desired angles of each quadrotor respect to its body frame which ${\psi }_d$ is zero; finally, $\dot{\varphi }, \dot{\theta }$ and $\dot{\psi }$ present the angular velocity of each quadrotor respect to its body frame. Moreover, in this paper, the desired angular velocities of each quadrotor ${\dot{\varphi }}_d, {\dot{\theta }}_d$ and ${\dot{\psi }}_d$ are supposed to be zero. At the end, the outputs of Eq. \ref{GrindEQ__78}-\ref{GrindEQ__80}, $U_{2i},U_{3i},and U_{4i}$ (i=1,...,n), are the one of inputs for the actuators, and $\tau_i\in\mathbb{R}^3$ in Eq. \ref{GrindEQ__22} is one of the output, and since we don’t study the dynamics of actuators, they are equal in this study, i.e. $\tau_i=[U_{2i},U_{3i}, U_{4i}]^T$. It must be mentioned that the other input and output of the actuators are, respectively, $U_{1i}$ which is derived from $\boldsymbol{U}_i$ in Eq. \ref{GrindEQ__74} and $f_i$.

\section{ Simulation}

In this section, numerical examples are presented to demonstrate the appropriate operation of the proposed control framework. First, the transportation of a homogeneous rigid body is studied, and then it is done for the transportation of a non-homogeneous load. In both cases, 4 quadrotors ($n=4$) are considered and each cable is divided into 5 sections of mass, spring, and damper ($n_i=5$). Parameters of the system and initial conditions are given in Table~\ref{tab:table1} where the parameters for the quadrotors and the cables are taken from the actual models in \cite{16pp} and \cite{20pp}. Indeed, the cable parameters are defined according to the material properties. Control parameters are presented in Table~\ref{tab:table2}.

\begin{table}
\caption{\label{tab:table1}Physical parameters of the system}\centering
\begin{tabular}{lccc} \hline 
Parameter  & Value & Unit \\ \hline  
$m_l$ & 2 &(kg) \\   
$r_{l0}$ & ${\left[0,0,0\right]}^T$ &(m) \\   
$v_{l0}$ & ${\left[0,0,0\right]}^T$ &(m/s) \\   
${\eta }_{l0}$ & ${\left[30,20,-30\right]}^{\mathrm{T}}$ &(degree) \\   
$\Omega_{l0}$ & ${\left[0,0,0\right]}^T$ &(rad/s) \\   
$m_{ij}$ & 0.01 &(kg) \\   
$K_{ij}$ & ${\mathrm{2\times 10}}^{\mathrm{4}}$ &(N/m) \\   
$b_{ij}$ & $20$ &(N.s/m) \\   
$L_{ij}$ & 0.5 &(m) \\   
$J_i$ & $\mathrm{diag}\left[\mathrm{0.577,0.577,1.05}\right]\mathrm{\times }{\mathrm{10}}^{\mathrm{-}\mathrm{2}}$ &(kg/m$ ^{2}$) \\   
$m_i$ & 0.755 &(kg) \\   
${\eta }_{0i}$ & ${\left[0,0,0\right]}^T$ &(degree) \\   
$\Omega_{0i}$ & ${\left[0,0,0\right]}^T$ &(rad/s) \\   
$r_{desired}$ & $[10,10,-10]$ &(m) \\   
$v_{desired}$ & ${\left[0,0,0\right]}^T$ &(m/s) \\ \hline 
\end{tabular}
\end{table}

\begin{table}
\caption{\label{tab:table2}Control parameters}
\centering
\begin{tabular}{lcccc} \hline 
Parameter  &  Value & Parameter  &  Value  \\ \hline 
 $K_p$ & $\mathrm{diag}\,[0.7,0.7,1]$ &$K$ & $\mathrm{diag}\,[10,10,0.05]$   \\   
$K_i$  & $\mathrm{diag}\,[0.07,0.07,0.1]$ &${\phi }_i$ & $\mathrm{diag}\,[8,8,8]$    \\   
$K_v$  & $\mathrm{diag}\,[0.7,0.7,1]$ &$k_d$ & $2.08$    \\   
$\mathit{\Lambda}$ & $\mathrm{diag}\,[0.05,0.05,0.0005]$ &$k_p$ & $0.2$  \\   
$M$  & $\mathrm{diag}\,[12,12,0.12]$  &$k_i$ & $0.1$ \\ \hline 
\end{tabular}
\end{table}

\subsection{Homogeneous Rigid Body}
In the first case, the load has a cuboid shape with the length, width and height of  3.2, 2.4 and 0.2 meters, respectively. The moments of inertia and connection points of cables are expressed in Tables~\ref{tab:table3}. The desired location is $\left[ \begin{array}{ccc}
10 & 10 & -10 \end{array}
\right]^T$.

\begin{table}
\caption{\label{tab:table3}The moments of inertia  of the homogeneous load and the connection points of cables}
\centering
\begin{tabular}{lccc} \hline 
Parameter  & Value & Unit \\ \hline 
$J_l$ & $\mathrm{diag[} \begin{array}{ccc}
\mathrm{1.45} & \mathrm{2.57} & \mathrm{4} \end{array}
\mathrm{]}$ &  (kg/m${}^{2}$) \\ 
$d_1$ & ${\left[\mathrm{1.2,-1.6,-0.2}\right]}^{\mathrm{T}}$  & (m) \\   
$d_2$ & ${\left[\mathrm{1.2,1.6,-0.2}\right]}^T$ & (m) \\   
$d_3$ & ${\left[\mathrm{-}\mathrm{1.2,-1.6,-0.2}\right]}^T$ & (m) \\   
$d_4$ & ${\left[\mathrm{-}\mathrm{1.2,1.6,-0.2}\right]}^{\mathrm{T}}$ & (m) \\ \hline 
\end{tabular}
\end{table}

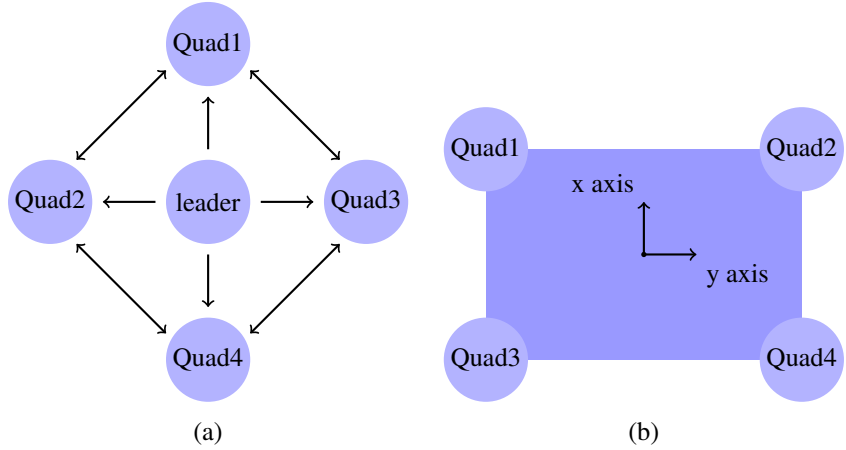
\begin{figure}
	\centering
	\begin{tabular}{c c}
	\begin{tikzpicture}[scale=0.7]
\fill[blue!30!white] (0,0) circle (0.8cm) node[black] {leader};
\fill[blue!30!white] (0,-3) circle (0.8cm) node[black] {Quad4};
\fill[blue!30!white] (-3,0) circle (0.8cm) node[black] {Quad2};
\fill[blue!30!white] (0,3) circle (0.8cm) node[black] {Quad1};
\fill[blue!30!white] (3,0) circle (0.8cm) node[black] {Quad3};
\draw[thick,->] (0,1) -- (0,2);
\draw[thick,->] (0,-1) -- (0,-2);
\draw[thick,->] (1,0) -- (2,0);
\draw[thick,->] (-1,0) -- (-2,0);
\draw[thick,<->] (2.5,0.8) -- (0.8,2.5);
\draw[thick,<->] (-2.5,0.8) -- (-0.8,2.5);
\draw[thick,<->] (2.5,-0.8) -- (0.8,-2.5);
\draw[thick,<->] (-2.5,-0.8) -- (-0.8,-2.5);
\end{tikzpicture} & \begin{tikzpicture}[scale=0.7]
\fill[blue!40!white] (0,0) rectangle (6,4);
\fill[blue!30!white] (0,0) circle (0.8cm) node[black] {Quad3};
\fill[blue!30!white] (0,4) circle (0.8cm) node[black] {Quad1};
\fill[blue!30!white] (6,0) circle (0.8cm) node[black] {Quad4};
\fill[blue!30!white] (6,4) circle (0.8cm) node[black] {Quad2};
\draw[thick,->] (3,2) -- (4,2) node[anchor=north west] {y axis};
\draw[thick,->] (3,2) -- (3,3) node[anchor=south east] {x axis};
\fill[black] (3,2) circle (0.05cm);
\end{tikzpicture} \\ (a) & (b) \\
	\end{tabular}
\caption{(a) The communication graph for the system (b) the quadrotors position respect to the rectangular load in X-Y plane }
\label{fig3}\end{figure}

With respect to dimensions of the load and connection points, the communication graph of the multi-agent transportation system and the quadrotors position respect to the rectangular load in the X-Y plane based on this communication graph are depicted in Figure~\ref{fig3}. The distance of the agents from each other and the leader in the direction $z$ is zero and in the directions of $x$ and $y$ are given in Table~\ref{tab:table4}. 

\begin{table}[hbt!]
\caption{\label{tab:table4}Distances between the  agents}
\centering
\begin{tabular}{lccc} \hline 
Parameter & x direction (m) &  y direction (m) \\ \hline 
$d_{1leader}$ &  1.2 &  -1.6 \\   
$d_{2leader}$ & 1.2 &  1.6 \\   
$d_{3leader}$ &  -1.2 &  -1.6 \\   
$d_{4leader}$ &  -1.2 &  1.6 \\   
\quad$d_{12}$ & 2.4 & 0 \\   
\quad$d_{13}$ & 0 &  3.2 \\   
\quad$d_{24}$ & 2.4 & 0 \\   
\quad$d_{34}$ &  0 &  3.2 \\ \hline 
\end{tabular}
\end{table}

\begin{figure}
\centering
\begin{tabular}{c c}
\includegraphics*[width=2.8in]{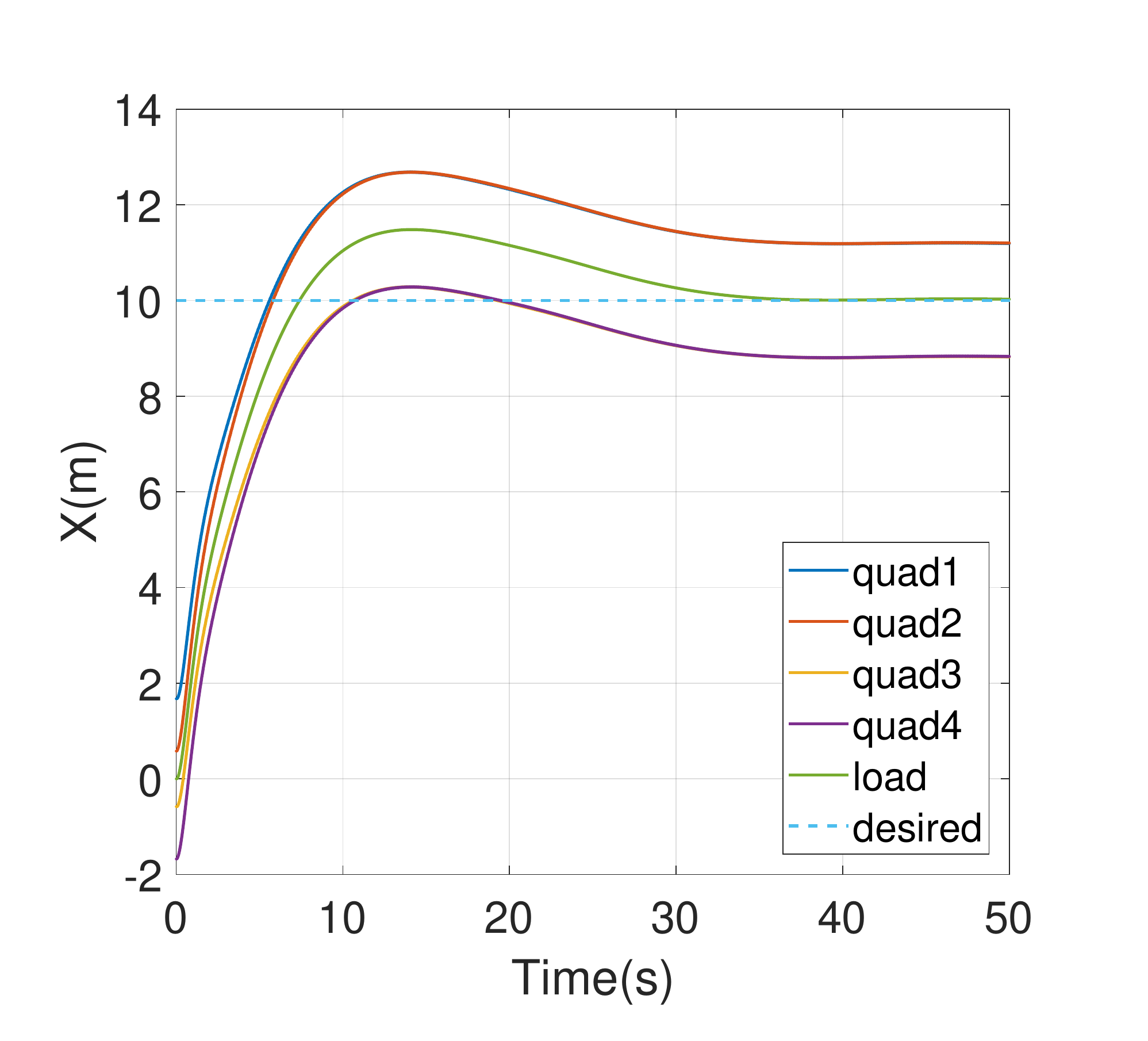} & \hspace*{-0.4in} \includegraphics*[width=2.8in]{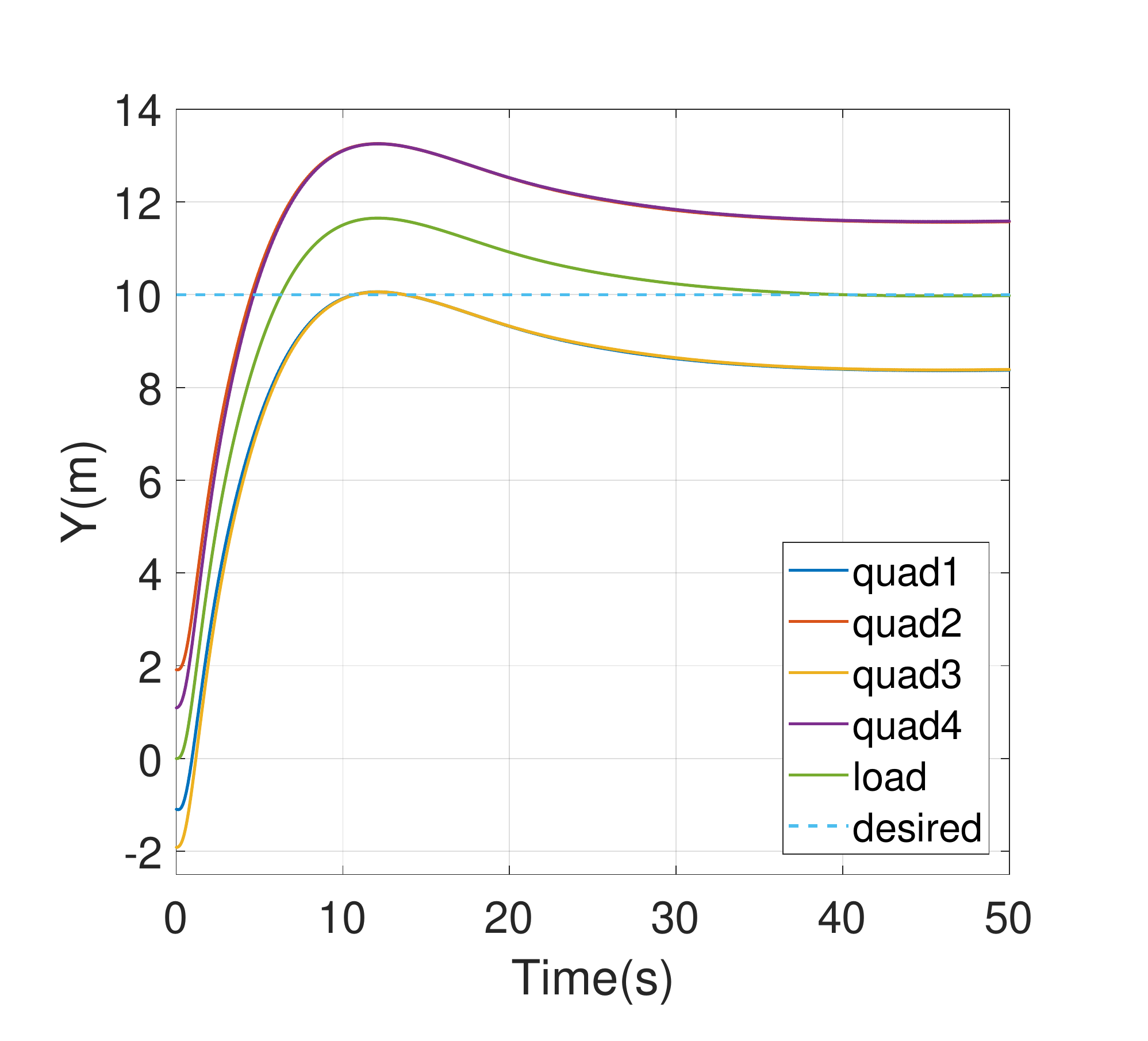} \\ (a) & (b) \\ \includegraphics*[width=2.8in]{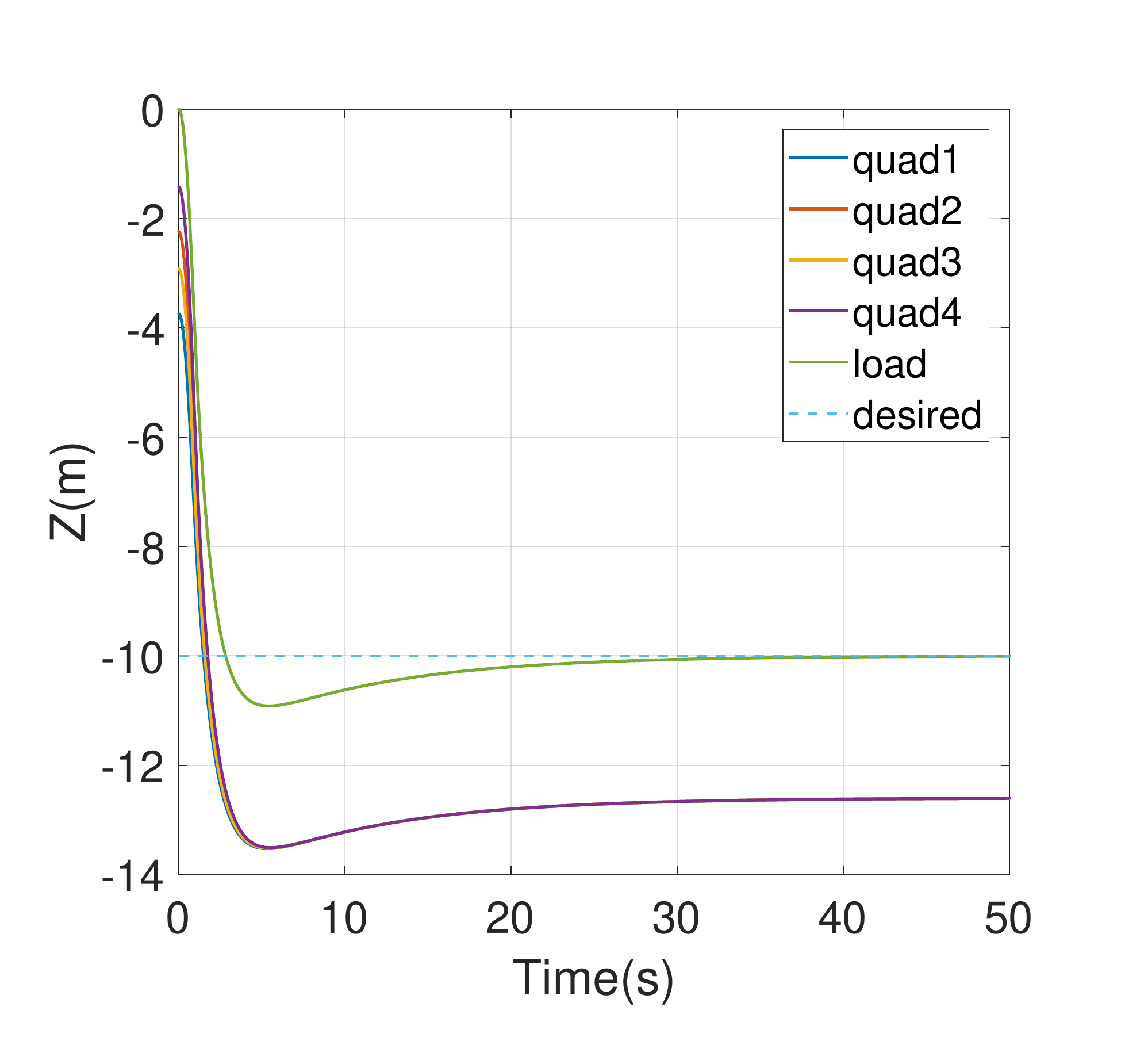} & \hspace*{-0.4in} \includegraphics*[width=2.8in]{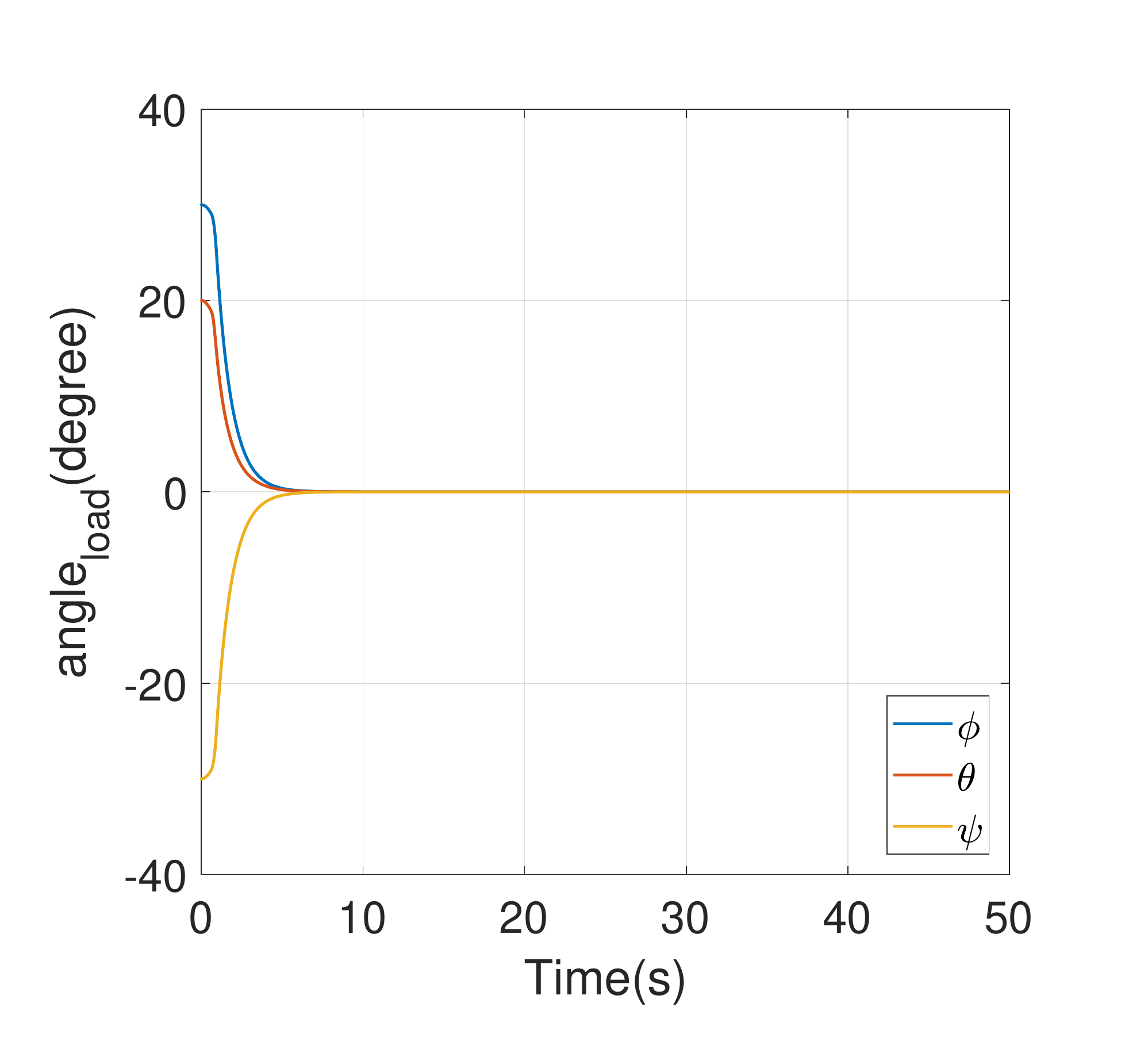}  \\ (c) & (d) \\
\includegraphics*[width=2.8in]{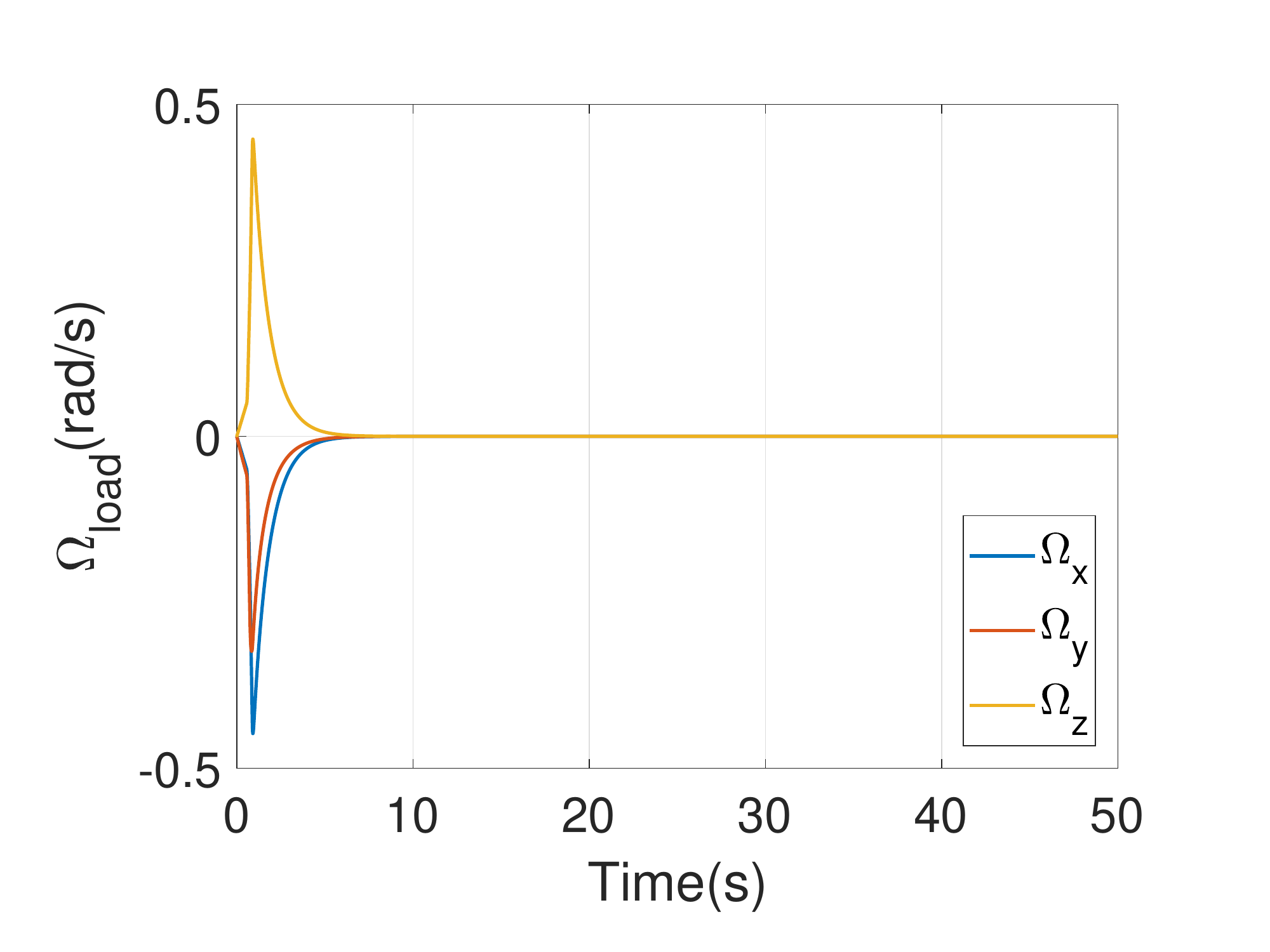} & \hspace*{-0.4in} \includegraphics*[width=2.8in]{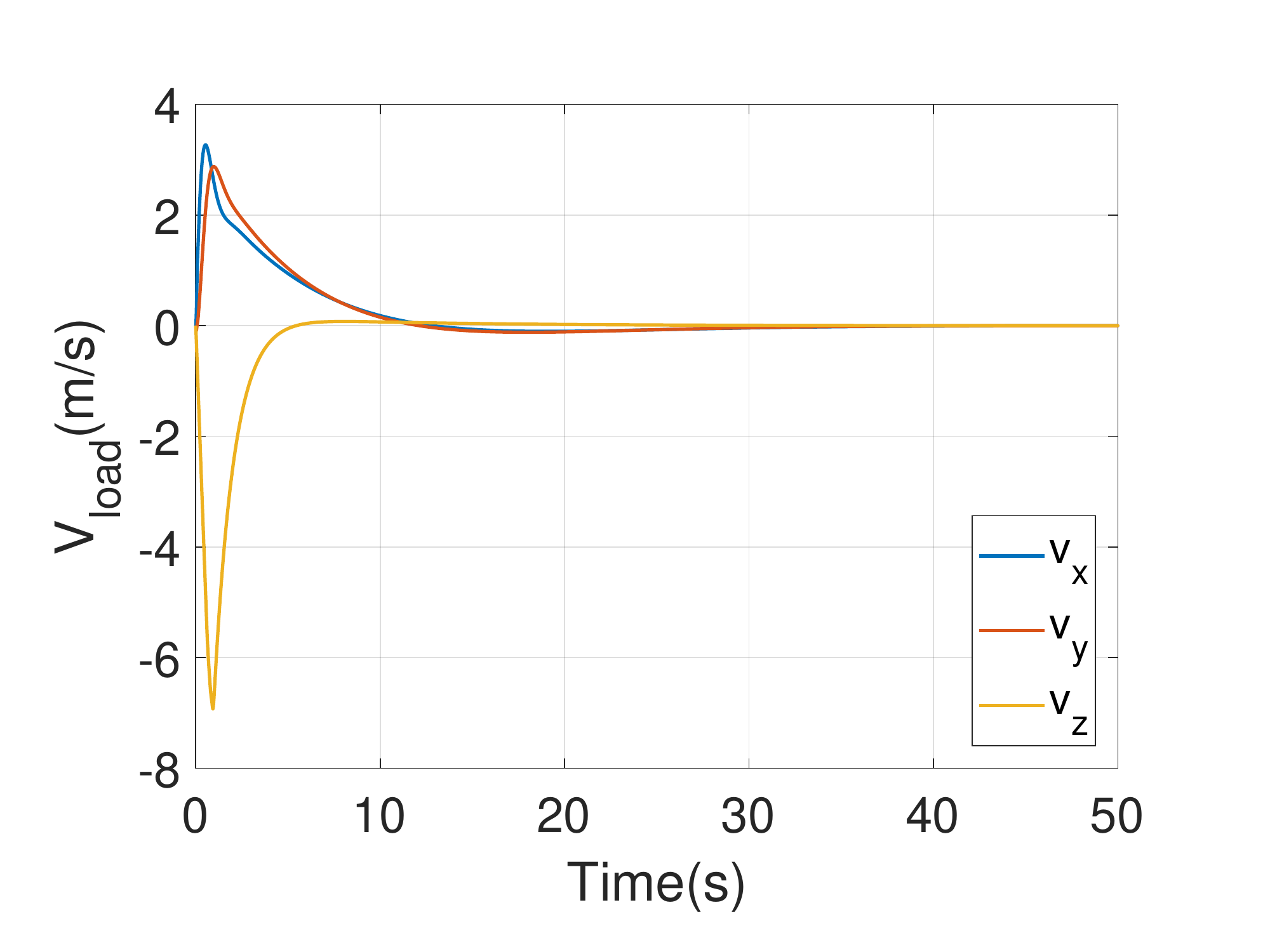} \\ (e) & (f) \\
\end{tabular}
\caption{Simulation results for transportation of a homogeneous rigid body: (a) position in $x$ direction, (b) position in $y$ direction, (c) position in $z$ direction, (d) load angles, (e) load angular velocity, (f) load velocity, (g) angular velocities of the first cable components, and (h) length change of the first cable.}\label{fig4}
\end{figure}

\begin{figure}
\centering
\begin{tabular}{c}
\includegraphics*[width=3.5in]{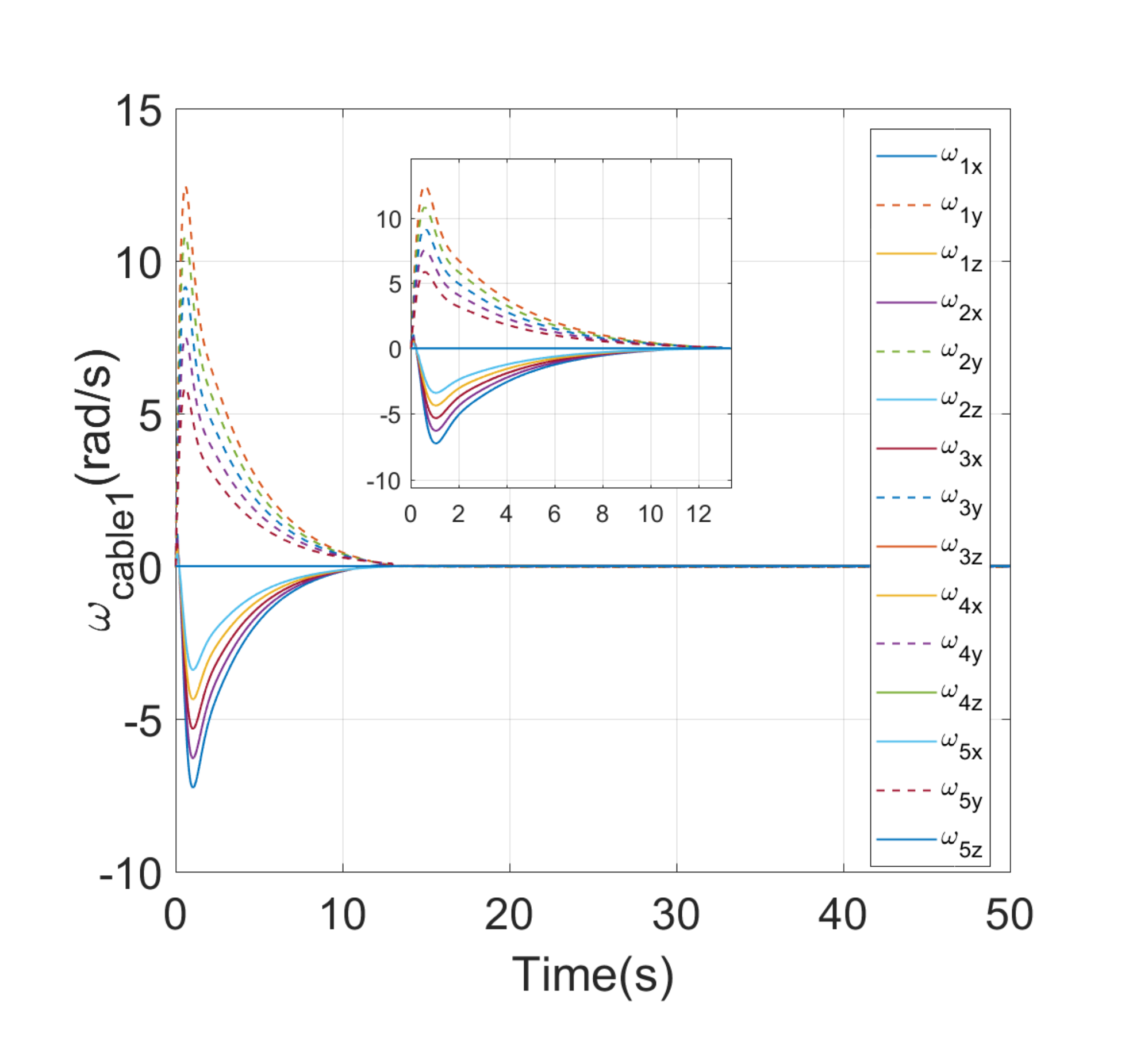} \\ (a) \\ \includegraphics*[width=3.5in]{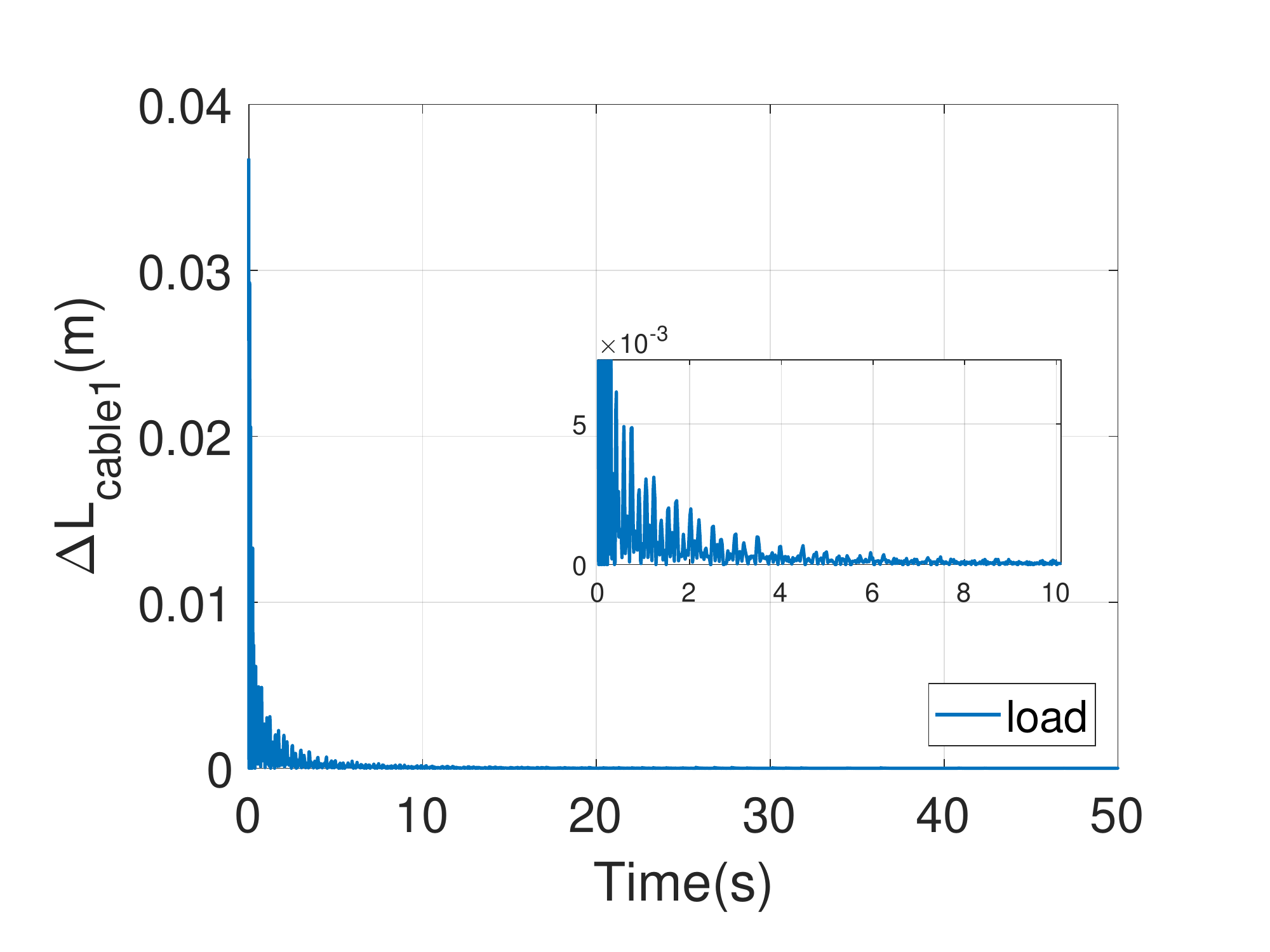}  \\ (b) \\
\end{tabular}
\caption{Simulation results for transportation of a homogeneous rigid body: (a) angular velocities of cable 1 components, and (b) length change of the connected cable 1.}\label{fig5}
\end{figure}

Figure~\ref{fig4} shows the simulation results. As it can be observed, although quadrotors are in different positions, after a while, they reach their desired formation and maintain it. In addition, the load is transported to the desired position by an appropriate velocity. Although, cables have some fluctuations at the initial moments (Figure~\ref{fig5}), as the quadrotors reach their final formation, these fluctuations are damped.

\subsection{ Non-homogeneous Rigid Body}
We have a cone with a radius of 1~m and a height of 3~m as a non-homogeneous load. According to the dimensions of the rigid body, the connection points and momentum of inertia are determined and reported in  Table~\ref{tab:table5}.

\begin{table}
\caption{\label{tab:table5}The moments of inertia of the non-homogeneous load and connection points of cables}
\centering
\begin{tabular}{lcc} \hline 
Parameter & Value  & Unit \\ \hline 
$J_l$ & $\mathrm{diag([} \begin{array}{ccc}
\mathrm{0.6} & \mathrm{2.1} & 2.1 \end{array}
\mathrm{])}$ & (kg/m)${}^{2}$ \\   
$d_1$ & ${\left[\mathrm{2.25,0,0}\right]}^{\mathrm{T}}$ & (m) \\   
$d_2$ & ${\left[\mathrm{0,-0.75,0}\right]}^{\mathrm{T}}$ & (m) \\   
$d_3$ & ${\left[\mathrm{0,0.75,0}\right]}^{\mathrm{T}}$ & (m) \\   
$d_4$ & ${\left[\mathrm{2.25,0.75,0}\right]}^{\mathrm{T}}$ & (m) \\   \hline 
\end{tabular}
\end{table}

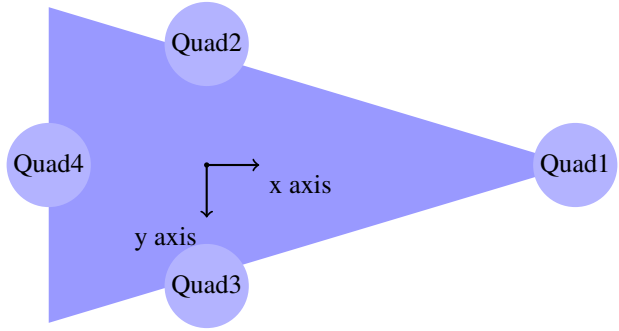
\begin{figure}
\centering
\begin{tikzpicture}[scale=0.7]
\fill[blue!40!white] (0,0) node[anchor=north]{}
  -- (0,6) node[anchor=north]{}
  -- (10,3) node[anchor=south]{}
  -- cycle;
  \fill[blue!30!white] (0,3) circle (0.8cm) node[black] {Quad4};
\fill[blue!30!white] (3,5.3) circle (0.8cm) node[black] {Quad2};
\fill[blue!30!white] (3,0.7) circle (0.8cm) node[black] {Quad3};
\fill[blue!30!white] (10,3) circle (0.8cm) node[black] {Quad1};
\draw[thick,->] (3,3) -- (4,3) node[anchor=north west] {x axis};
\draw[thick,->] (3,3) -- (3,2) node[anchor=north east] {y axis};
\fill[black] (3,3) circle (0.05cm);
\end{tikzpicture}
\caption{The quadrotors position respect to the cone load in X-Y plane}
\label{fig6}
\end{figure}

As the mass distribution is not homogeneous, the quadrotors must be located properly based on the mass distribution. Therefore, the communication graph is the same for the homogeneous load (Figure~\ref{fig4}(a)), but the quadrotors positions respect to the nonhomogeneous load (the cone) in the X-Y plane are defined as Figure~\ref{fig6} where most of quadrotors are assigned to the part of the cone which has more mass distribution. The distances are given in Table~\ref{tab:table6}.

\begin{table}
\caption{\label{tab:table6}Distances between the agents}
\centering
\begin{tabular}{lcc} \hline 
Parameter &x direction (m) & y direction (m) \\ \hline 
$d_{1leader}$ & 0 & 2.25 \\   
$d_{2leader}$ & -0.75 & 0 \\   
$d_{3leader}$ & 0.75 & 0 \\   
$d_{4leader}$ & 0 & -0.75 \\   
$d_{12}$ & 0.75 & 2.25 \\   
$d_{13}$ & 0.75 & 2.25 \\   
$d_{24}$ & 0.75 & 0.75 \\   
$d_{34}$ & 0.75 & 0.75 \\ \hline 
\end{tabular}
\end{table}

\begin{figure}
\centering
\begin{tabular}{c c}
\includegraphics*[width=2.8in]{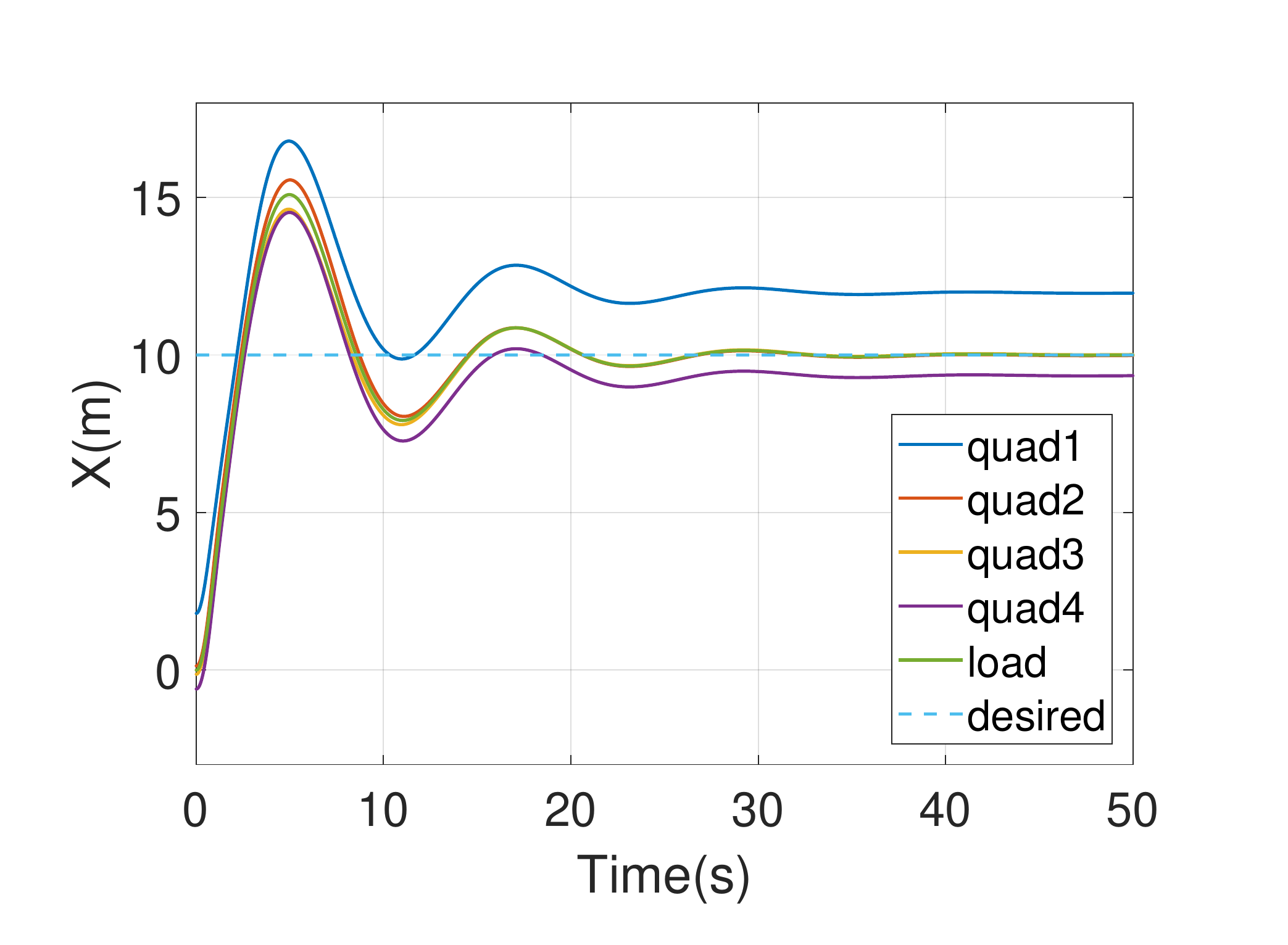} & \hspace*{-0.4in}
\includegraphics*[width=2.8in]{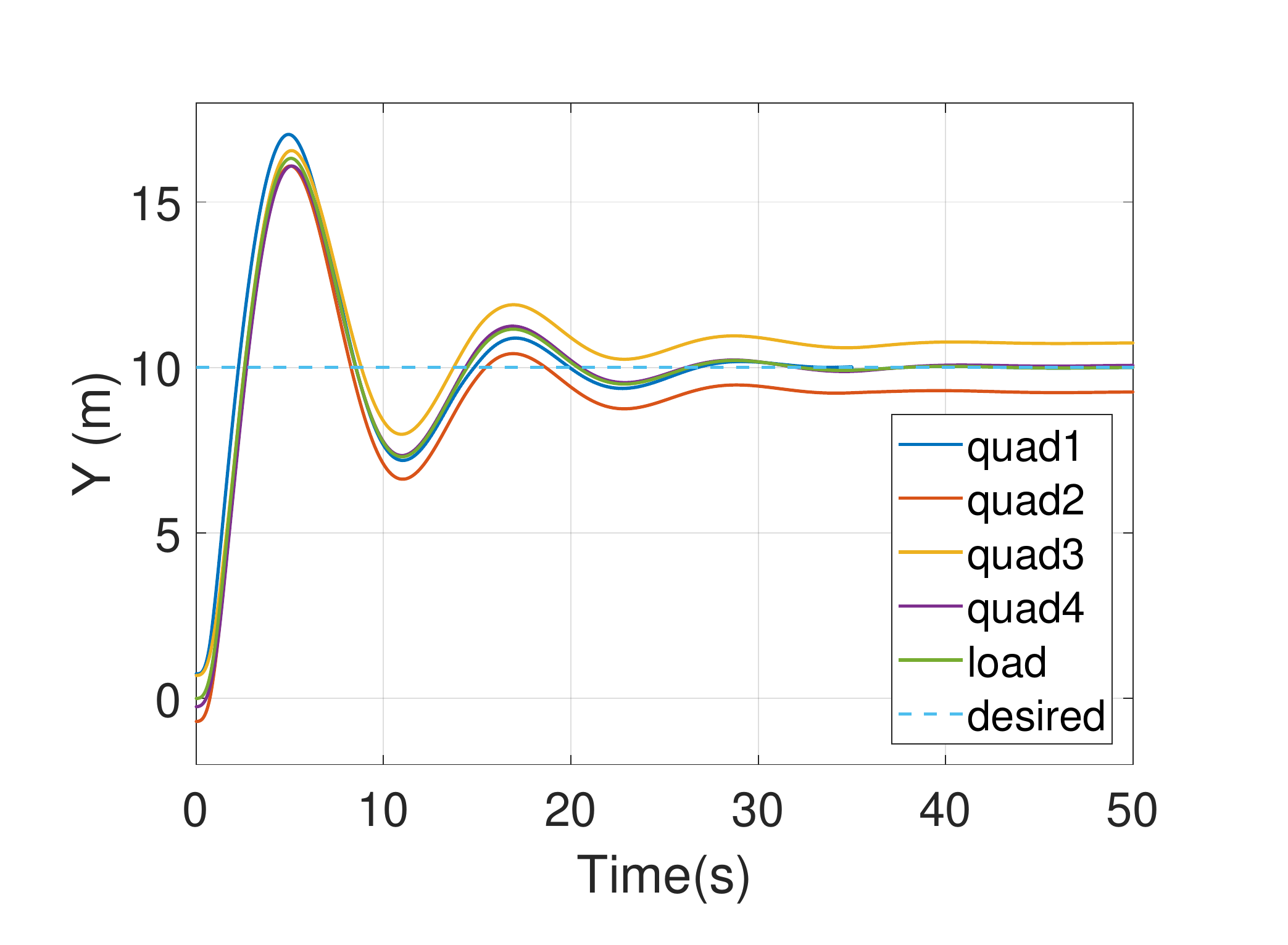} \\ (a) & (b) \\
\includegraphics*[width=2.8in]{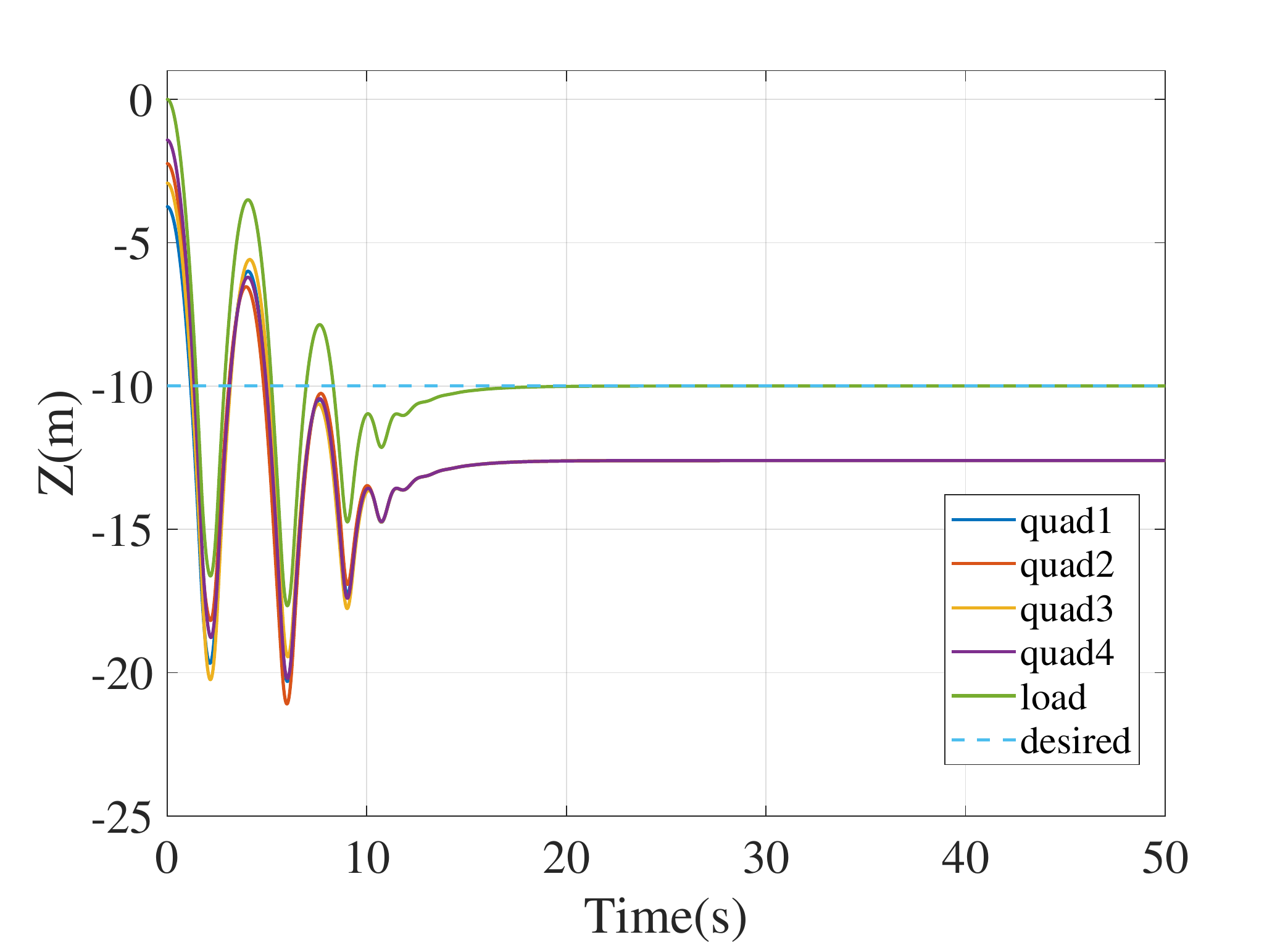} & \hspace*{-0.4in}
\includegraphics*[width=2.8in]{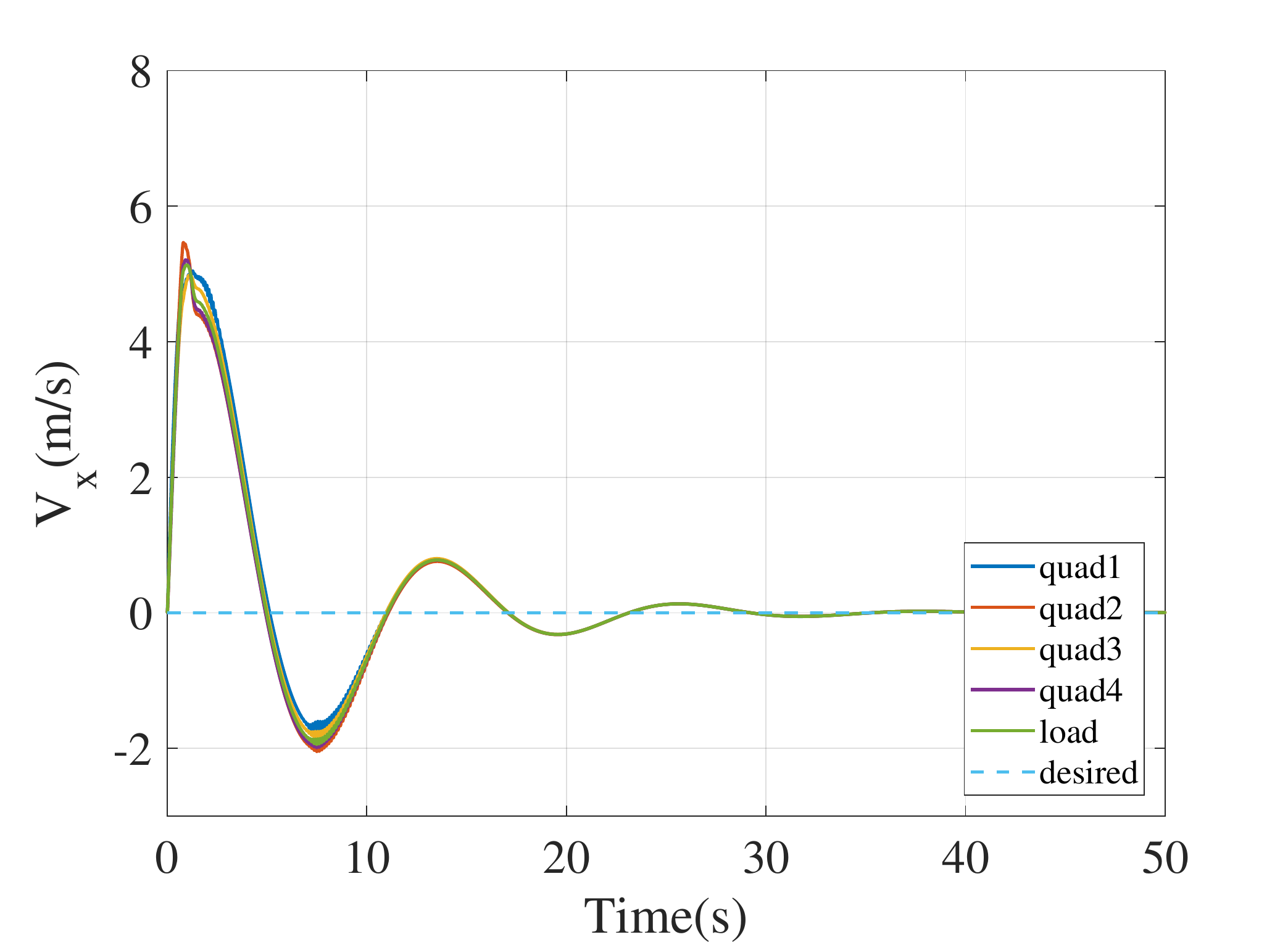} \\ (c) & (d) \\
\includegraphics*[width=2.8in]{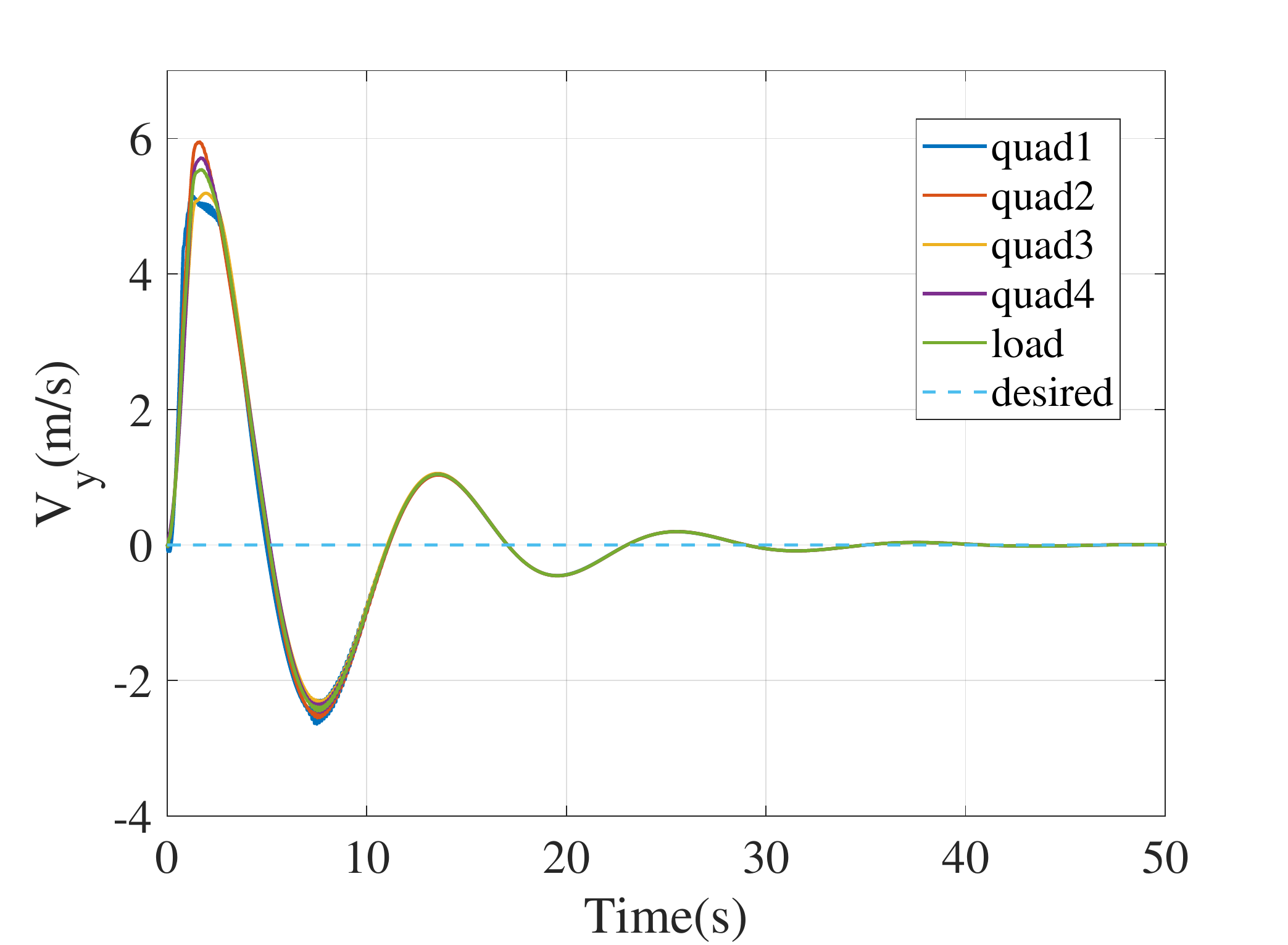} & \hspace*{-0.4in}
\includegraphics*[width=2.8in]{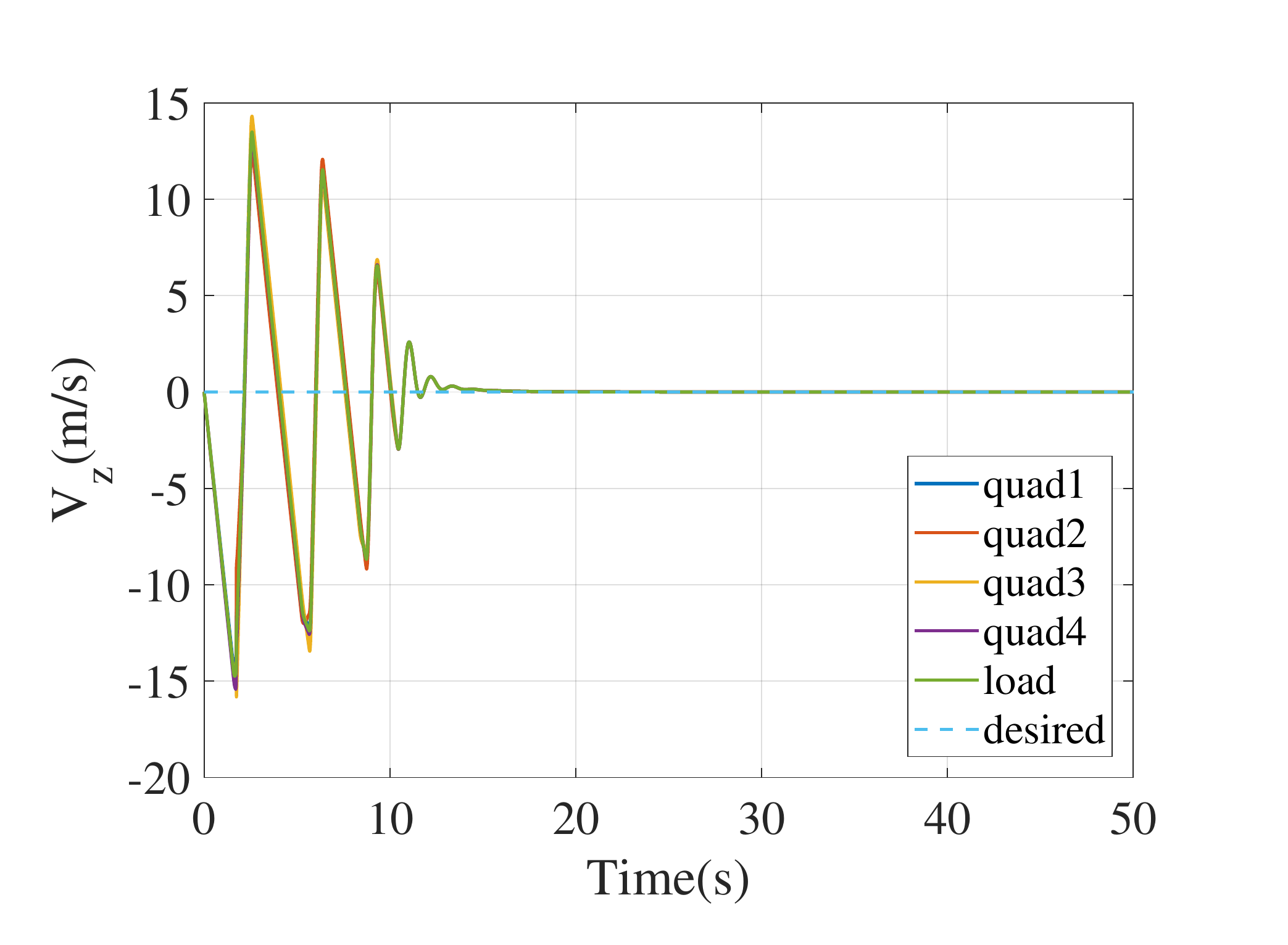} \\ (e) & (f) \\
\end{tabular}
\caption{Simulation results of the transportation of non-homogeneous rigid body: (a) position in $x$ direction, (b) position in $y$ direction, (c) position in $z$ direction, (d) velocity in $x$ direction, (e) velocity in $y$ direction, and (f) velocity in $z$ direction.}
\label{fig7}
\end{figure}

The numerical result of aerial transportation of a non-homogeneous slung rigid body is presented in Figure~\ref{fig7}. As seen, the transportation system of the non-homogeneous rigid body reaches the desired formation in about 20 seconds (Figure~\ref{fig7}(b)) that is larger than 10 seconds for the homogeneous payload in  (Figure~\ref{fig4}(b)). Moreover, transportation of the non-homogeneous rigid body has more swings and fluctuations in comparison with the homogeneous load. Hence, the non-uniform distribution of the mass highly influences the transportation. However, there are some oscillations, but one of the advantages of the formation control of a multi-agent system, which arranges quadrotors in accordance with the mass distribution, minimizes load oscillations and prevents quadrotor collisions during transport. 

\begin{figure}
\centering
\begin{tabular}{c c}
\includegraphics*[width=2.8in]{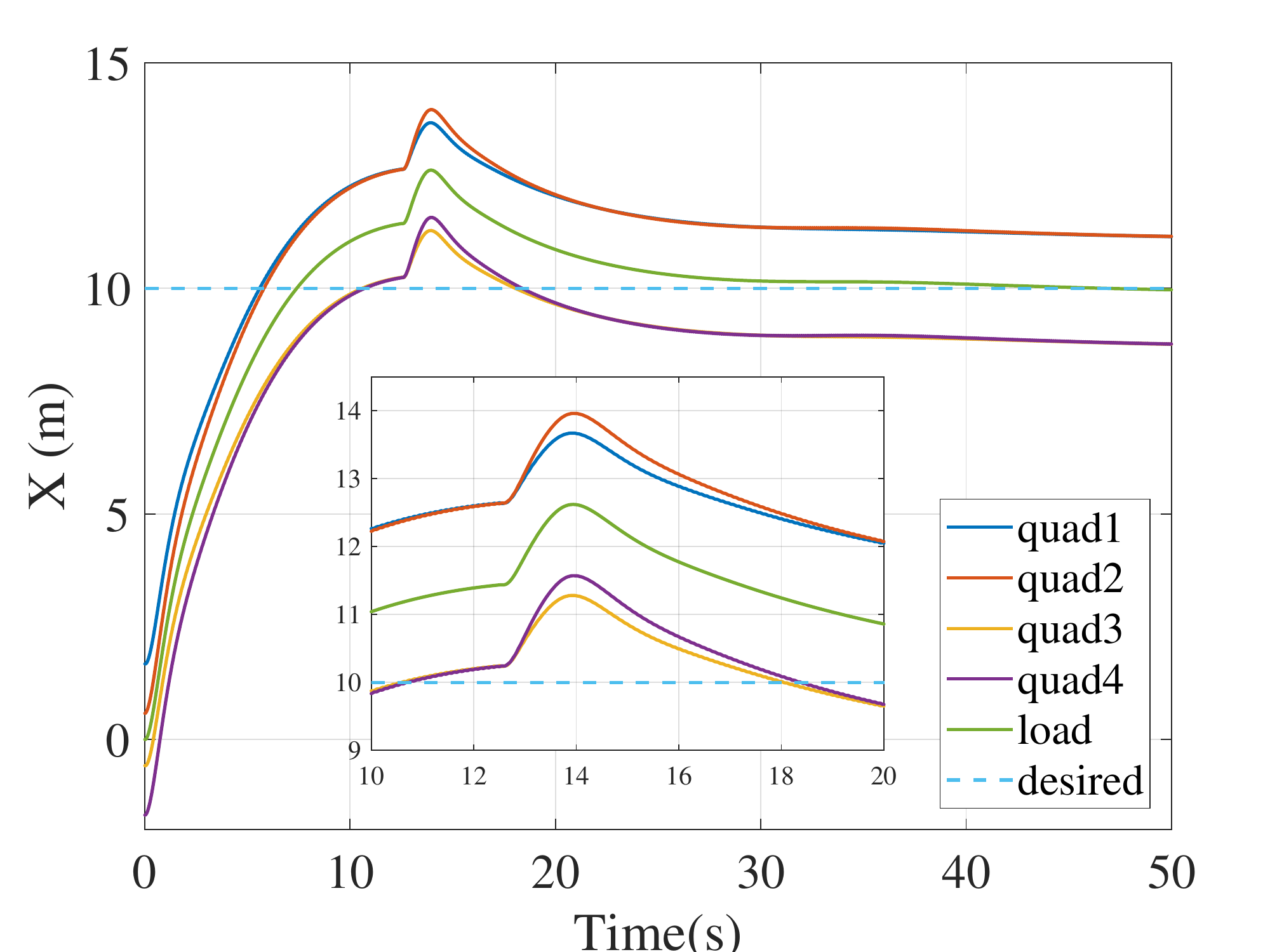} & \hspace*{-0.4in} \includegraphics*[width=2.8in]{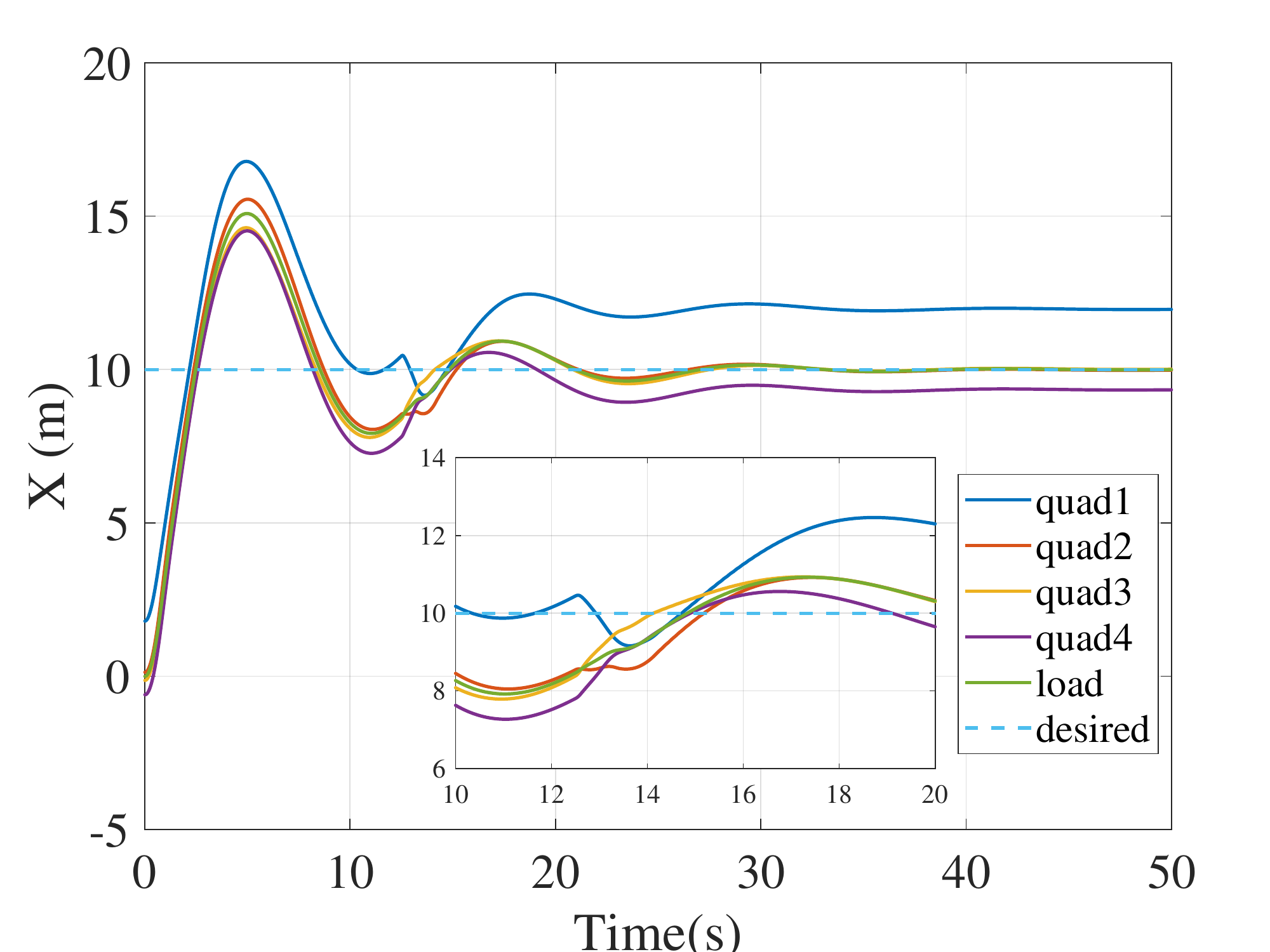} \\ (a) & (b) \\
\includegraphics*[width=3in]{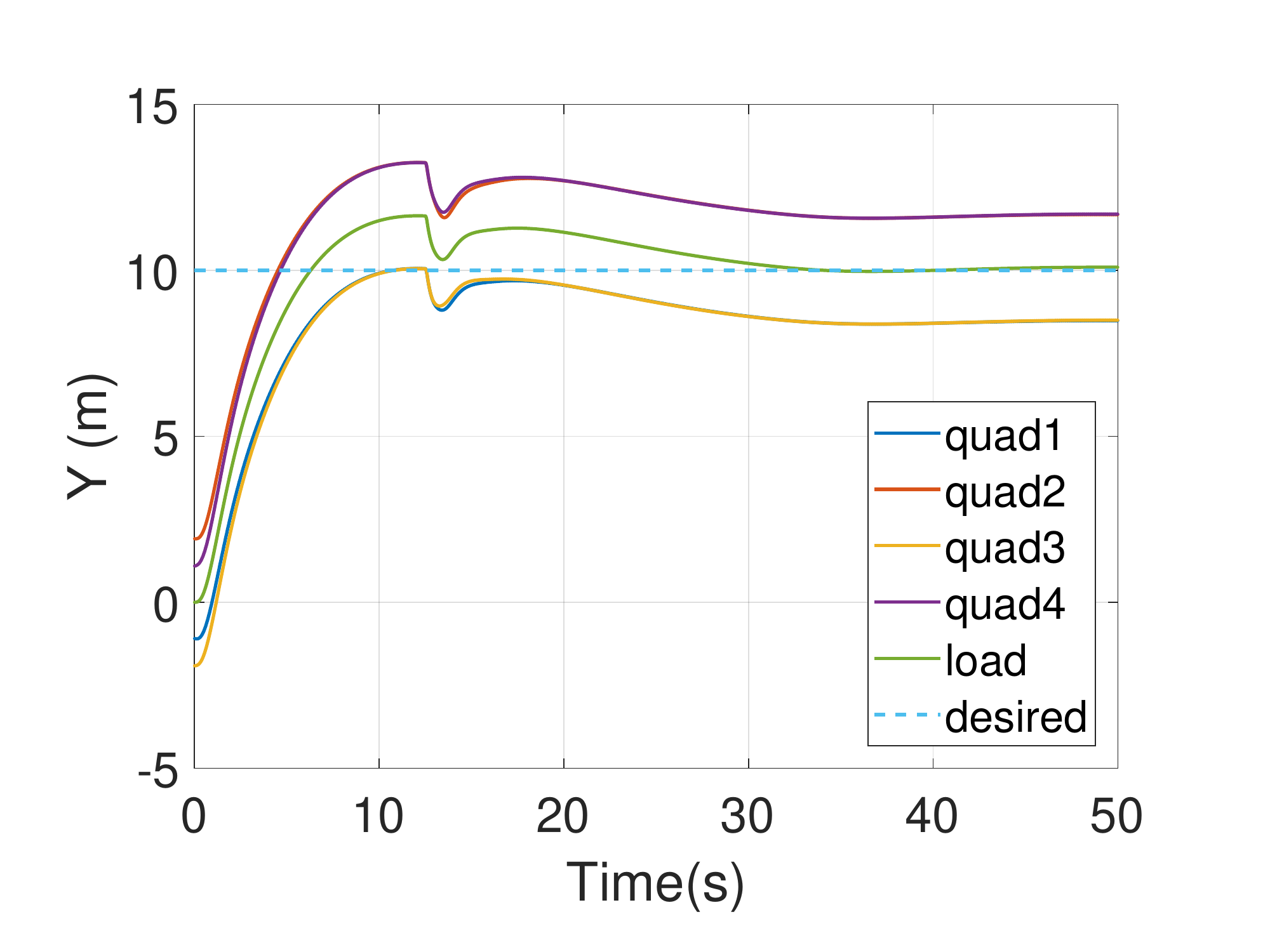} & \hspace*{-0.4in} \includegraphics*[width=3in]{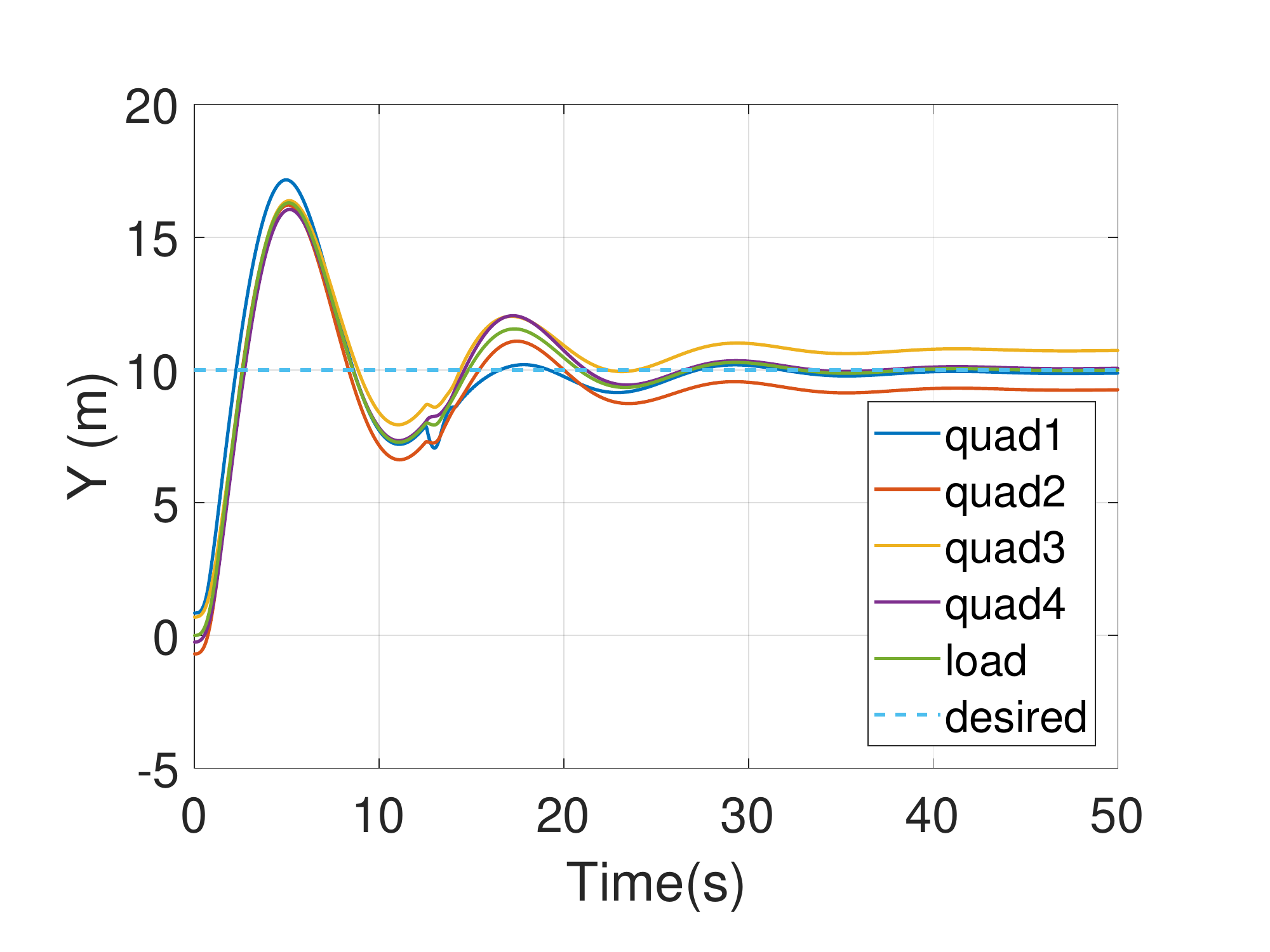}  \\ (c) & (d) \\
\end{tabular}
\caption{Aerial load transportation in presence of disturbances: (a) position in $x$ direction of homogeneous, (a) position in $x$ direction of nonhomogeneous, (c) position in $y$ direction of homogeneous and (d) position in $y$ direction of homogeneous.}\label{fig8}
\end{figure}

In order to indicate that the designed control strategy is robust against disturbances, the aerial load transportation is examined in presence of impulse disturbance which is exerted on quadrotors 1 and 3 at 12.5 seconds (Figure~\ref{fig8}). As it is expected, the transportation system for carrying the homogeneous payload confronts less perturbation (Figure~\ref{fig8}(a)) in comparison with the transportation system of nonhomogeneous body (Figure~\ref{fig8}(b)).

\section{ Conclusions }
In this paper, aerial transportation of a non-uniform rigid body, which was connected through elastic and flexible cables to a group of quadrotors, was investigated. The Euler-Lagrange method was used to derive the motion equations of the system. Since in real world the objects mostly do not have a uniform and homogeneous shape and mass distribution,  the payload was assumed as 6 DOF  non-uniform rigid body and each cable was considered as a combination of masses, springs, and dampers. Then, a control strategy (consisting of navigation control, formation control, and attitude control) was employed for multi-agent transportation of the rigid body to the desired position with the desired velocity as well as put the formation of quadrotors according to the payload mass distribution. Navigation control was designed based on the length of cables, payload position and velocity, and its errors. The sliding mode approach was applied to maintain the formation control which was able to set quadrotors according to the mass distribution, reduce load swings and avoid collision of quadrotors during transporting. Furthermore, the asymptotic stability was proven for the designed formation control. PID control was used to stabilize the attitudes of each quadrotor. Finally, the capability of the proposed control strategy for conveying homogeneous and non-homogeneous bodies was shown by numerical simulations, where the quadrotors were spread out around the load based on its mass distribution. According to the numerical results,  the transportation system besides having an appropriate performance in conveying the uniform load, it was able to carry a payload with nonuniform mass distribution, however damping the load fluctuations of nonuniform load took more time.

\section*{Conflicts of interest}

The authors declare no conflicts of interest exist. A preprint of this paper is available at 	arXiv:2305.09175.

\appendix
\section{Appendix}
\subsection{Deriving the equations of motion}\label{app:a}
 By replacing \eqref{GrindEQ__4} into \eqref{GrindEQ__9},  and 
defining $\ M_T=m_l+\sum^n_{i=1}{(m_i}+\sum^{n_i}_{j=1}{m_{ij})}$, ${M_{qi}=m}_i+\sum^{n_i}_{j=1}{m_{ij}}$ and $M_{cij}=m_i+\sum^{j-1}_{a=1}{m_{ia}}$, the kinetic and potential energies can be rewritten as
\begin{equation}
\label{GrindEQ__12}
\begin{aligned}
T=&\frac{1}{2}M_T{\left\|\boldsymbol{\dot{r}_{l}}\right\|}^2-\frac{1}{2}\sum^n_{i=1}{M_{qi}{\boldsymbol\Omega_{l}}^{T}{\widehat{\boldsymbol{d}}}^{2}_{\boldsymbol{i}}{\boldsymbol\Omega}_{\boldsymbol{l}}~\ }+\frac{1}{2}{\boldsymbol\Omega_{l}}^{T}{\boldsymbol{J}}_{\boldsymbol{l}}{\boldsymbol\Omega}_{\boldsymbol{l}}+\frac{1}{2}\sum^n_{i=1}{{\boldsymbol\Omega_{i}}^{T}_{\boldsymbol{i}}{\boldsymbol{J}}_{\boldsymbol{i}}{\boldsymbol\Omega}_{\boldsymbol{i}}~\ }\\&+\sum^n_{i=1}{\sum^{n_i}_{j,k=1}{M_{cij}l_{ij}l_{ik}{\boldsymbol{\dot{q}}}^{T}_{\boldsymbol{ij}}{\boldsymbol{\dot{q}}}_{\boldsymbol{ik}}}}+\sum^n_{i=1}{\sum^{n_i}_{j,k=1}{M_{cij}{\dot{l}}_{ij}{\dot{l}}_{ik}{\boldsymbol{q}}^{T}_{\boldsymbol{ij}}{\boldsymbol{q}}_{\boldsymbol{ik}}}}+\sum^n_{i=1}{M_{qi}{\boldsymbol\Omega}^{T}_{\boldsymbol{l}}{\widehat{\boldsymbol{d}}}_{\boldsymbol{i}}{\boldsymbol{R}}^{T}_{\boldsymbol{l}}\boldsymbol{\dot{r}_{l}}} \\&-\sum^n_{i=1}{\sum^{n_i}_{j=1}{M_{cij}l_{ij}{\boldsymbol{\dot{r}}}^{T}_{\boldsymbol{l}}\boldsymbol{\dot{q}_{ij}}}}-\sum^n_{i=1}{\sum^{n_i}_{j=1}{M_{cij}{\dot{l}}_{ij}{\boldsymbol{\dot{r}}}^{T}_{\boldsymbol{l}}\boldsymbol{q_{ij}}}}-\sum^n_{i=1}{\sum^{n_i}_{j=1}{M_{cij}l_{ij}{\boldsymbol\Omega}^{T}_{\boldsymbol{l}}{\widehat{\boldsymbol{d}}}_{\boldsymbol{i}}{\boldsymbol{R}}^{T}_{\boldsymbol{l}}\boldsymbol{\dot{q}_{ij}}}} \\&-\sum^n_{i=1}{\sum^{n_i}_{j=1}{M_{cij}{\dot{l}}_{ij}{\boldsymbol\Omega}^{T}_{\boldsymbol{l}}{\widehat{\boldsymbol{d}}}_{\boldsymbol{i}}{\boldsymbol{R}}^{T}_{\boldsymbol{l}}\boldsymbol{q_{ij}}}}+\sum^n_{i=1}{\sum^{n_i}_{j,k=1 \atop
k\neq j}{M_{cij}l_{ij}{\dot{l}}_{ik}{\boldsymbol{\dot{q}}}^{T}_{\boldsymbol{ij}}{\boldsymbol{\dot{q}}}_{\boldsymbol{ik}}}}
\end{aligned}
\end{equation}
\begin{equation}\label{GrindEQ__13}\begin{aligned} 
V=&-M_Tg \boldsymbol{e_3}\cdot \boldsymbol{r_l}-\sum^n_{i=1}{\sum^{n_i}_{j=1}{M_{cij}gl_{ij}\boldsymbol{q_{ij}} \cdot \boldsymbol{e_3}}}-\sum^n_{i=1}{M_{qi}g\boldsymbol{R_l}\boldsymbol{d_i} \cdot \boldsymbol{e_3}} \\&+\sum^n_{i=1}{\sum^{n_i}_{j=1}{\frac{1}{2}K_{ij}\frac{\mathrm{\Delta }l_{ij}\left(\left|\mathrm{\Delta }l_{ij}\right|+\mathrm{\Delta }l_{ij}\right)}{4}}} 
\end{aligned}\end{equation}

Afterwards, the Lagrangian is determined by $L=T-V$. Furthermore, the effect of damping $F$ is determined as
\begin{equation}\label{GrindEQ__14}\begin{aligned} 
F=\sum^n_{i=1}{\sum^{n_i}_{j=1}{\frac{1}{2}b_{ij}{\dot{l}}^2_{ij}\frac{1+\sign (\Delta l_{ij})}{2}}} 
\end{aligned}\end{equation}

According to~\cite{9pp}, small variations of $\boldsymbol{R_i} \in \mathrm{\text{SO}(3)}$ in terms of exponential map is
\begin{equation}\label{GrindEQ__15}
\delta \boldsymbol{R_i}=\frac{d}{d\varepsilon }\boldsymbol{R_i}{\exp \left(\varepsilon {\widehat{\boldsymbol{\eta }}}_{\boldsymbol{i}}\right)\ }=\boldsymbol{R_i}{\widehat{\boldsymbol{\eta }}}_{\boldsymbol{i}} 
\end{equation}
where ${\boldsymbol{\eta }}_{\boldsymbol{i}}  \in\mathbb{R}^3$and the differential is computed at $\varepsilon =0$. In addition, the variations of angular velocity and $\boldsymbol{q_{ij}} \in S^{2}$ are given by 
\begin{align}\label{GrindEQ__16}
\delta {\boldsymbol\Omega}_{\boldsymbol{i}}&={\boldsymbol{\dot{\eta }}}_{\boldsymbol{i}} + {\boldsymbol\Omega}_{\boldsymbol{i}}\times{\boldsymbol{\eta }}_{\boldsymbol{i}} &
\delta \boldsymbol{q_{ij}} = {\boldsymbol{\xi }}_{\boldsymbol{ij}}\times\boldsymbol{q_{ij}} 
\end{align}
where ${\boldsymbol{\xi_{ij} }}  \in\mathbb{R}^3$ satisfies $\boldsymbol{\xi_{ij}} \cdot \boldsymbol{q_{ij}} = 0$. Finally, based on the Euler-Lagrange equation, the motion equations of the proposed aerial transportation of a rigid body are presented as 
\begin{equation}\label{GrindEQ__18}\begin{split} 
M_T{\boldsymbol{\ddot{r}}}_{\boldsymbol{l}}&+\sum^n_{i=1}{M_{qi}\boldsymbol{R_l}\left({\widehat{\boldsymbol\Omega}}^{2}_{\boldsymbol{l}}\boldsymbol{d_i} - {\widehat{\boldsymbol{d}}}_{\boldsymbol{i}}{\boldsymbol{\dot\Omega}}_{\boldsymbol{l}}\right)}-M_Tg \boldsymbol{e_3}-\sum^n_{i=1}{\sum^{n_i}_{j=1}{M_{cij}\left(l_{ij}{\boldsymbol{\ddot{q}}}_{\boldsymbol{ij}} + 2{\dot{l}}_{ij}\boldsymbol{\dot{q}_{ij}} + {\ddot{l}}_{ij}\boldsymbol{q_{ij}}\right)}}=-\sum^n_{i=1}{f_i\boldsymbol{R_i} \boldsymbol{e_3}}
\end{split}\end{equation}

\begin{equation}\label{GrindEQ__19}\begin{split} 
&\left({\boldsymbol{J}}_{\boldsymbol{l}} -\sum^n_{i=1}{M_{qi}}{\widehat{\boldsymbol{d}}}^{2}_{\boldsymbol{i}}\right){\boldsymbol{\dot\Omega}}_{\boldsymbol{l}}+\sum^n_{i=1}{M_{qi}}{\widehat{\boldsymbol{d}}}_{\boldsymbol{i}}{\boldsymbol{R}}^{T}_{\boldsymbol{l}}{\boldsymbol{\ddot{r}}}_{\boldsymbol{l}} -\sum^n_{i=1}{\sum^{n_i}_{j=1}{M_{cij}{\widehat{\boldsymbol{d}}}_{\boldsymbol{i}}{\boldsymbol{R}}^{T}_{\boldsymbol{l}}\left(l_{ij}{\boldsymbol{\ddot{q}}}_{\boldsymbol{ij}}+2{\dot{l}}_{ij}\boldsymbol{\dot{q}_{ij}}+{\ddot{l}}_{ij}\boldsymbol{q_{ij}}\right)}} \\ &\quad +{\widehat{\boldsymbol\Omega}}_{\boldsymbol{l}}\left({\boldsymbol{J}}_{\boldsymbol{l}}-\sum^n_{i=1}{M_{qi}}{\widehat{\boldsymbol{d}}}^{2}_{\boldsymbol{i}}\right){\boldsymbol\Omega}_{\boldsymbol{l}} - \sum^n_{i=1}{M_{qi}}g{\widehat{\boldsymbol{d}}}_{\boldsymbol{i}}{\boldsymbol{R}}^{T}_{\boldsymbol{l}} \boldsymbol{e_3}=-\sum^n_{i=1}{{f_i{\widehat{\boldsymbol{d}}}_{\boldsymbol{i}}{\boldsymbol{R}}^{T}_{\boldsymbol{l}}\boldsymbol{R_i}\boldsymbol{e_3} }}
\end{split}\end{equation}

\begin{equation}\label{GrindEQ__20}\begin{split} 
& \sum^{n_i}_{k=1}{M_{cij}l_{ij}l_{ik}{\widehat{\boldsymbol{q}}}^{2}_{\boldsymbol{ij}}{\boldsymbol{\ddot{q}}}_{\boldsymbol{ik}}}+2\sum^{n_i}_{k=1}{M_{cij}l_{ij}{\dot{l}}_{ik}{\widehat{\boldsymbol{q}}}^{2}_{\boldsymbol{ij}}{\boldsymbol{\dot{q}}}_{\boldsymbol{ik}}}+\sum^{n_i}_{k=1 \atop k\neq j}{M_{cij}l_{ij}{\ddot{l}}_{ik}{\widehat{\boldsymbol{q}}}^{2}_{\boldsymbol{ij}}{\boldsymbol{q}}_{\boldsymbol{ik}}}  -M_{cij}{\dot{l}}^2_{ij}{\widehat{\boldsymbol{q}}}^{2}_{\boldsymbol{ij}}\boldsymbol{q_{ij}}\\
& \quad-M_{cij}l_{ij}{\widehat{\boldsymbol{q}}}^{2}_{\boldsymbol{ij}}\left({\boldsymbol{\ddot{r}}}_{\boldsymbol{l}} + \boldsymbol{R_l}\left({\widehat{\boldsymbol\Omega}}^{2}_{\boldsymbol{l}}\boldsymbol{d_i} - {\widehat{\boldsymbol{d}}}_{\boldsymbol{i}}{\boldsymbol{\dot\Omega}}_{\boldsymbol{l}}\right)\right)+M_{cij}gl_{ij}{\widehat{\boldsymbol{q}}}^{2}_{\boldsymbol{ij}} \boldsymbol{e_3}=l_{ij}{f_i{\widehat{\boldsymbol{q}}}^{2}_{\boldsymbol{ij}}\boldsymbol{R_i}\boldsymbol{e_3} }
\end{split}\end{equation}

\begin{equation}\label{GrindEQ__21}\begin{split} 
&  \sum^{n_i}_{k=1}{M_{cij}{\ddot{l}}_{ik}{\boldsymbol{q}}^{T}_{\boldsymbol{ij}}{\boldsymbol{q}}_{\boldsymbol{ik}}}+\sum^{n_i}_{k=1 \atop k\neq j }{M_{cij}{\boldsymbol{q}}^{T}_{\boldsymbol{ij}}\left(2{\dot{l}}_{ik}{\boldsymbol{q}}_{\boldsymbol{ik}} + l_{ik}{\boldsymbol{\ddot{q}}}_{\boldsymbol{ik}}\right)} -M_{cij}{\boldsymbol{q}}^{T}_{\boldsymbol{ij}}\left({\boldsymbol{\ddot{r}}}_{\boldsymbol{l}} + \boldsymbol{R_l}\left({\widehat{\boldsymbol\Omega}}^{T}_{\boldsymbol{l}}\boldsymbol{d_i} - {\widehat{\boldsymbol{d}}}_{\boldsymbol{i}}{\boldsymbol{\dot\Omega}}_{\boldsymbol{l}}\right)\right) \\
&\quad-M_{cij}l_{ij}{\boldsymbol{\dot{q}}}^{T}_{\boldsymbol{ij}}\boldsymbol{\dot{q}_{ij}}+M_{cij}g{\boldsymbol{q}}^{T}_{\boldsymbol{ij}} \boldsymbol{e_3}+K_{ij}\mathit{\Delta}l_{ij}\frac{1+\sign(\mathit{\Delta}l_{ij})}{2}+b_{ij}{\dot{l}}_{ij}\frac{1+\sign(\mathit{\Delta}l_{ij})}{2}=-{f_i{\boldsymbol{q}}^{T}_{\boldsymbol{ij}}\boldsymbol{R_i}\boldsymbol{e_3} }
\end{split}\end{equation}

In the motion equation \eqref{GrindEQ__21}, $\frac{1+\sign\mathrm{(}\mathrm{\Delta }l_{ij})}{2}\ $in the force terms of springs and dampers is considered to model the behavior of cables during stretching and being loose. In other words, if the springs and dampers are compressed, they specify that the cable is loose because there is no tensile and damping forces as $\mathrm{\Delta }l_{ij}<0$. And, during the stretch of the springs and dampers ($\mathrm{\Delta }l_{ij}>0$), the tensile and damping forces are corresponding to $K_{ij}\mathrm{\Delta }L_{ij}$ and $b_{ij}{\dot{l}}_{ij}$, respectively.

Since $\boldsymbol{q_{ij}}\cdot \boldsymbol{{\dot{q}}_{ij}}=0$, the term $\boldsymbol{{\widehat{q}}}^2_{\boldsymbol{ij}}\boldsymbol{{\ddot{q}}_{ij}}$ can be rewritten as


\begin{equation}\label{GrindEQ__23}\begin{split} 
{\widehat{\boldsymbol{q}}}^{2}_{\boldsymbol{ij}}{\boldsymbol{\ddot{q}}}_{\boldsymbol{ij}} = \left(\boldsymbol{q_{ij}}\boldsymbol{\cdot }{\boldsymbol{\ddot{q}}}_{\boldsymbol{ij}}\right)\boldsymbol{q_{ij}} - \left(\boldsymbol{q_{ij}}\boldsymbol{\cdot }\boldsymbol{q_{ij}}\right){\boldsymbol{\ddot{q}}}_{\boldsymbol{ij}}
= -\left(\boldsymbol{\dot{q}_{ij}}\boldsymbol{\cdot }\boldsymbol{\dot{q}_{ij}}\right)\boldsymbol{q_{ij}} - \left(\boldsymbol{q_{ij}}\boldsymbol{\cdot }\boldsymbol{q_{ij}}\right){\boldsymbol{\ddot{q}}}_{\boldsymbol{ij}}
=  - \boldsymbol{\parallel }\boldsymbol{\dot{q}_{ij}}{\boldsymbol{\parallel }}^{2}\boldsymbol{q_{ij}} - {\boldsymbol{\ddot{q}}}_{\boldsymbol{ij}} 
\end{split}\end{equation}

Thus, the first term of \eqref{GrindEQ__20} can be represented by 
\begin{equation}\label{GrindEQ__24}
\sum^{n_i}_{k=1}{M_{cij}l_{ij}l_{ik}{\widehat{\boldsymbol{q}}}^{2}_{\boldsymbol{ij}}{\boldsymbol{\ddot{q}}}_{\boldsymbol{ik}}}=\ \sum^{n_i}_{k=1 \atop k\neq j}{M_{cij}l_{ij}l_{ik}{\widehat{\boldsymbol{q}}}^{2}_{\boldsymbol{ij}}{\boldsymbol{\ddot{q}}}_{\boldsymbol{ik}}}-M_{cij}l^2_{ij}{\boldsymbol{\ddot{q}}}_{\boldsymbol{ij}}-M_{cij}l^2_{ij}\parallel \boldsymbol{\dot{q}_{ij}}{\parallel }^2\boldsymbol{q_{ij}}
\end{equation}

Differentiating \eqref{GrindEQ__7} and considering that $\boldsymbol{q_{ij}}\cdot\boldsymbol{\omega_{ij}} = \boldsymbol{0}$, we have
\begin{equation}\label{GrindEQ__25}\begin{split} 
{\boldsymbol{\ddot{q}}}_{\boldsymbol{ij}} &= {\boldsymbol{\dot{\omega }}}_{\boldsymbol{ij}}\times\boldsymbol{q_{ij}} + \boldsymbol{\omega }\times\left(\boldsymbol{\omega_{ij}}\times\boldsymbol{q_{ij}}\right) = {\boldsymbol{\dot{\omega }}}_{\boldsymbol{ij}}\times\boldsymbol{q_{ij}} - \boldsymbol{\parallel }\boldsymbol{\omega_{ij}}{\boldsymbol{\parallel }}^{2}\boldsymbol{q_{ij}} \\
&= \textbf{\textit{\newline }} - {\widehat{\boldsymbol{q}}}_{\boldsymbol{ij}}{\boldsymbol{\dot{\omega }}}_{\boldsymbol{ij}} - \boldsymbol{\parallel }\boldsymbol{\omega_{ij}}{\boldsymbol{\parallel }}^{2}\boldsymbol{q_{ij}}.
\end{split}\end{equation}

Substituting \eqref{GrindEQ__18} into \eqref{GrindEQ__21} yields the equation of motions \eqref{GrindEQ__26}--\eqref{GrindEQ__22}.

\subsection{Construct the matrices}\label{app:b}
The mass matrix is defined by

\begin{equation}\label{GrindEQ__31}\begin{aligned} 
\boldsymbol{M}=&\left[ \begin{array}{cccccccccc}
{\boldsymbol{M}}_{r_l} & {\boldsymbol{M}}_{r_l{\eta }_l} & {\boldsymbol{M}}_{r_l{\omega }_{1j}} & {\boldsymbol{M}}_{r_l{\omega }_{2j}} & \cdots  & {\boldsymbol{M}}_{r_l{\omega }_{nj}} & {\boldsymbol{M}}_{r_ll_{1j}} &{\boldsymbol{M}}_{r_ll_{2j}} & \cdots  & {\boldsymbol{M}}_{r_ll_{nj}}\\ 
{\boldsymbol{M}}_{{\eta }_lr_l} & {\boldsymbol{J}}_{{\eta }_l} & {\boldsymbol{M}}_{{\eta }_l{\omega }_{1j}} & {\boldsymbol{M}}_{{\eta }_l{\omega }_{2j}} & \cdots  & {\boldsymbol{M}}_{{\eta }_l{\omega }_{nj}} & {\boldsymbol{M}}_{{\eta }_ll_{1j}}& {\boldsymbol{M}}_{{\eta }_ll_{2j}} & \cdots  & {\boldsymbol{M}}_{{\eta }_ll_{nj}} \\ 
{\boldsymbol{M}}_{{\omega }_{1j}r_l} & {\boldsymbol{M}}_{{\omega }_{1j}{\eta }_l} & {\boldsymbol{M}}_{{\omega }_{1j}} & \boldsymbol{0} & \cdots  & \boldsymbol{0} & {\boldsymbol{M}}_{{\omega }_{1j}l_{1j}} & \boldsymbol{0} & \cdots  & \boldsymbol{0} \\ 
{\boldsymbol{M}}_{{\omega }_{2j}r_l} & {\boldsymbol{M}}_{{\omega }_{2j}{\eta }_l} & \boldsymbol{0} & {\boldsymbol{M}}_{{\omega }_{2j}} & \cdots  & \boldsymbol{0} & \boldsymbol{0}  & {\boldsymbol{M}}_{{\omega }_{2j}l_{2j}} & \cdots  & \boldsymbol{0} \\ 
\vdots  & \vdots  & \vdots  & \vdots  & \ddots  & \vdots  & \vdots    & \vdots  & \ddots  & \vdots\\ 
{\boldsymbol{M}}_{{\omega }_{nj}r_l\ } & {\boldsymbol{M}}_{{\omega }_{nj}\eta _l} & \boldsymbol{0} & \boldsymbol{0} & \cdots  & {\boldsymbol{M}}_{{\omega }_{nj}} & \boldsymbol{0} & \boldsymbol{0} & \cdots  & {\boldsymbol{M}}_{{\omega }_{nj}l_{nj}}\\ 
{\boldsymbol{M}}_{l_{1j}r_l} & {\boldsymbol{M}}_{l_{1j}{\eta }_l} & {\boldsymbol{M}}_{l_{1j}{\omega }_{1j}} & \boldsymbol{0} & \cdots  & \boldsymbol{0} & {\boldsymbol{M}}_{l_{1j}}& \boldsymbol{0} & \cdots  & \boldsymbol{0} \\ 
{\boldsymbol{M}}_{l_{2j}r_l} & {\boldsymbol{M}}_{l_{2j}{\eta }_l} & \boldsymbol{0} & {\boldsymbol{M}}_{l_{2j}{\omega }_{2j}} & \cdots  & \boldsymbol{0} & \boldsymbol{0}   & {\boldsymbol{M}}_{l_{2j}} & \cdots  & \boldsymbol{0}\\ 
\vdots  & \vdots  & \vdots  & \vdots  & \ddots  & \vdots  & \vdots   & \vdots  & \ddots  & \vdots \\ 
{\boldsymbol{M}}_{l_{nj}r_l} & {\boldsymbol{M}}_{l_{nj}{\eta }_l} & \boldsymbol{0} & \boldsymbol{0} & \cdots  & {\boldsymbol{M}}_{l_{nj}{\omega }_{nj}} & \boldsymbol{0}& \boldsymbol{0} & \cdots  & {\boldsymbol{M}}_{l_{nj}}
\end{array}
\right]  
\end{aligned}\end{equation}
where sub-matrices are 
\begin{equation}\label{GrindEQ__32}\begin{aligned} 
{\boldsymbol{M}}_{r_l}=M_T{\boldsymbol{I}}_3
\end{aligned}\end{equation}
\begin{equation}\label{GrindEQ__33}\begin{aligned}
{\boldsymbol{M}}_{r_l{\eta }_l}=-\sum^n_{i=1}{M_{qi}\boldsymbol{R_l}{\widehat{\boldsymbol{d}}}_{\boldsymbol{i}}{\boldsymbol{\dot\Omega}}_{\boldsymbol{l}}}
\end{aligned}\end{equation}
\begin{equation}
{\boldsymbol{M}}_{{\eta }_lr_l}={\boldsymbol{M}}^T_{r_l{\eta }_l}
\end{equation}
\begin{equation}\label{GrindEQ__34}\begin{aligned} 
{\boldsymbol{M}}_{r_l{\omega }_{ij}}=[ \begin{array}{cccc}
M_{ci1}l_{i1}{\widehat{\boldsymbol{q}}}_{\boldsymbol{i}\boldsymbol{1}} & M_{ci2}l_{i2}{\widehat{\boldsymbol{q}}}_{\boldsymbol{i}\boldsymbol{2}} & \cdots  & M_{cim}l_{im}{\widehat{\boldsymbol{q}}}_{\boldsymbol{im}} \end{array}
]
\end{aligned}\end{equation}
\begin{equation}\label{GrindEQ__35}\begin{aligned} 
{\boldsymbol{M}}_{r_ll_{ij}}=-[ \begin{array}{cccc}
M_{ci1}{\boldsymbol{q}}_{\boldsymbol{i}\boldsymbol{1}} & M_{ci2}{\boldsymbol{q}}_{\boldsymbol{i}\boldsymbol{2}} & \cdots  & M_{cim}{\boldsymbol{q}}_{\boldsymbol{im}} \end{array}
]
\end{aligned}\end{equation}
\begin{equation}\label{GrindEQ__36}\begin{aligned}
{\boldsymbol{J}}_{{\eta }_l}=\left({\boldsymbol{J}}_{\boldsymbol{l}}-\sum^n_{i=1}{M_{qi}}{\widehat{\boldsymbol{d}}}^2_{\boldsymbol{i}}\right){\boldsymbol{\dot\Omega}}_{\boldsymbol{l}}
\end{aligned}\end{equation}
\begin{equation}\label{GrindEQ__37}\begin{aligned} 
{\boldsymbol{M}}_{{\eta }_l{\omega }_{ij}}=[ \begin{array}{cccc}
M_{ci1}l_{i1}{\widehat{\boldsymbol{d}}}_{\boldsymbol{i}}{\boldsymbol{R}}^{T}_{\boldsymbol{l}}{\widehat{\boldsymbol{q}}}_{\boldsymbol{i}\boldsymbol{1}} & M_{ci2}l_{i2}{\widehat{\boldsymbol{d}}}_{\boldsymbol{i}}{\boldsymbol{R}}^{T}_{\boldsymbol{l}}{\widehat{\boldsymbol{q}}}_{\boldsymbol{i}\boldsymbol{2}} & \cdots  & M_{cim}l_{im}{\widehat{\boldsymbol{d}}}_{\boldsymbol{i}}{\boldsymbol{R}}^{T}_{\boldsymbol{l}}{\widehat{\boldsymbol{q}}}_{\boldsymbol{im}} \end{array}
]
\end{aligned}\end{equation}
\begin{equation}\label{GrindEQ__38}\begin{aligned} 
{\boldsymbol{M}}_{{\eta }_ll_{ij}}=-[ \begin{array}{cccc}
M_{ci1}{\widehat{\boldsymbol{d}}}_{\boldsymbol{i}}{\boldsymbol{R}}^{T}_{\boldsymbol{l}}{\boldsymbol{q}}_{\boldsymbol{i}\boldsymbol{1}} & M_{ci2}{\widehat{\boldsymbol{d}}}_{\boldsymbol{i}}{\boldsymbol{R}}^{T}_{\boldsymbol{l}}{\boldsymbol{q}}_{\boldsymbol{i}\boldsymbol{2}} & \cdots  & M_{cim}{\widehat{\boldsymbol{d}}}_{\boldsymbol{i}}{\boldsymbol{R}}^{T}_{\boldsymbol{l}}{\boldsymbol{q}}_{\boldsymbol{im}} \end{array}
]
\end{aligned}\end{equation}
\begin{equation}\label{GrindEQ__39}\begin{aligned} 
{\boldsymbol{M}}_{{\omega }_{ij}r_l}=-{\left[ \begin{array}{cccc}
M_{ci1}l_{i1}{\widehat{\boldsymbol{q}}}^{2}_{\boldsymbol{i}\boldsymbol{1}} & M_{ci2}l_{i2}{\widehat{\boldsymbol{q}}}^{2}_{\boldsymbol{i}\boldsymbol{2}} & \cdots  & M_{cim}l_{im}{\widehat{\boldsymbol{q}}}^{2}_{\boldsymbol{im}} \end{array}
\right]}^T
\end{aligned}\end{equation}
\begin{equation}\label{GrindEQ__40}\begin{aligned} 
{\boldsymbol{M}}_{{\omega }_{ij}{\eta }_l}={\left[ \begin{array}{cccc}
M_{ci1}l_{i1}{\widehat{\boldsymbol{q}}}^{2}_{\boldsymbol{i}\boldsymbol{1}}\boldsymbol{R_l}{\widehat{\boldsymbol{d}}}_{\boldsymbol{i}} & M_{ci2}l_{i2}{\widehat{\boldsymbol{q}}}^{2}_{\boldsymbol{i}\boldsymbol{2}}\boldsymbol{R_l}{\widehat{\boldsymbol{d}}}_{\boldsymbol{i}} & \cdots  & M_{cim}l_{im}{\widehat{\boldsymbol{q}}}^{2}_{\boldsymbol{im}}\boldsymbol{R_l}{\widehat{\boldsymbol{d}}}_{\boldsymbol{i}} \end{array}
\right]}^T
\end{aligned}\end{equation}
\begin{equation}\label{GrindEQ__41}\begin{aligned} 
{\boldsymbol{M}}_{l_{ij}r_l}=-{\left[ \begin{array}{cccc}
M_{ci1}{\boldsymbol{q}}^{T}_{\boldsymbol{i}\boldsymbol{1}} & M_{ci2}{\boldsymbol{q}}^{T}_{\boldsymbol{i}\boldsymbol{2}} & \cdots  & M_{cim}{\boldsymbol{q}}^{T}_{\boldsymbol{im}} \end{array}
\right]}^T
\end{aligned}\end{equation}
\begin{equation}\label{GrindEQ__42}\begin{aligned} 
{\boldsymbol{M}}_{l_{ij}{\eta }_l}={\left[ \begin{array}{cccc}
M_{ci1}{\boldsymbol{q}}^{T}_{\boldsymbol{i}\boldsymbol{1}}\boldsymbol{R_l}{\widehat{\boldsymbol{d}}}_{\boldsymbol{i}} & M_{ci2}{\boldsymbol{q}}^{T}_{\boldsymbol{i}\boldsymbol{2}}\boldsymbol{R_l}{\widehat{\boldsymbol{d}}}_{\boldsymbol{i}} & \cdots  & M_{cim}{\boldsymbol{q}}^{T}_{\boldsymbol{im}}\boldsymbol{R_l}{\widehat{\boldsymbol{d}}}_{\boldsymbol{i}} \end{array}
\right]}^T
\end{aligned}\end{equation}
\begin{equation}\label{GrindEQ__43}\begin{aligned} 
{\boldsymbol{M}}_{{\omega }_{ij}}=\left[ \begin{array}{cccc}
M_{ci1}l^2_{i1}{\widehat{\boldsymbol{q}}}_{\boldsymbol{i}\boldsymbol{1}} & M_{ci1}l_{i1}l_{i2}{\widehat{\boldsymbol{q}}}^{2}_{\boldsymbol{i}\boldsymbol{1}}{\widehat{\boldsymbol{q}}}_{\boldsymbol{i}\boldsymbol{2}} & \cdots  & M_{ci1}l_{i1}l_{im}{\widehat{\boldsymbol{q}}}^{2}_{\boldsymbol{i}\boldsymbol{1}}{\widehat{\boldsymbol{q}}}_{\boldsymbol{im}} \\ 
M_{ci2}l_{i2}l_{i1}{\widehat{\boldsymbol{q}}}^{2}_{\boldsymbol{i}\boldsymbol{2}}{\widehat{\boldsymbol{q}}}_{\boldsymbol{i}\boldsymbol{1}} & M_{ci2}l^2_{i2}{\widehat{\boldsymbol{q}}}_{\boldsymbol{i}\boldsymbol{2}} & \cdots  & M_{ci2}l_{i2}l_{im}{\widehat{\boldsymbol{q}}}^{2}_{\boldsymbol{i}\boldsymbol{2}}{\widehat{\boldsymbol{q}}}_{\boldsymbol{im}} \\ 
\vdots  & \vdots  & \ddots  & \vdots  \\ 
M_{cim}l_{im}l_{i1}{\widehat{\boldsymbol{q}}}^{2}_{\boldsymbol{im}}{\widehat{\boldsymbol{q}}}_{\boldsymbol{i}\boldsymbol{1}} & M_{cim}l_{im}l_{i2}{\widehat{\boldsymbol{q}}}^{2}_{\boldsymbol{im}}{\widehat{\boldsymbol{q}}}_{\boldsymbol{i}\boldsymbol{2}} & \cdots  & M_{cim}l^2_{im}{\widehat{\boldsymbol{q}}}_{\boldsymbol{im}} \end{array}
\right]
\end{aligned}\end{equation}
\begin{equation}\label{GrindEQ__44}\begin{aligned} 
{\boldsymbol{M}}_{l_{ij}}=\left[ \begin{array}{cccc}
M_{ci1} & M_{ci1}{\boldsymbol{q}}^{T}_{\boldsymbol{i}\boldsymbol{1}}{\boldsymbol{q}}_{\boldsymbol{i}\boldsymbol{2}} & \cdots  & M_{ci1}{\boldsymbol{q}}^{T}_{\boldsymbol{i}\boldsymbol{1}}{\boldsymbol{q}}_{\boldsymbol{im}} \\ 
M_{ci2}{\boldsymbol{q}}^{T}_{\boldsymbol{i}\boldsymbol{2}}{\boldsymbol{q}}_{\boldsymbol{i}\boldsymbol{1}} & M_{ci2} & \cdots  & M_{ci2}{\boldsymbol{q}}^{T}_{\boldsymbol{i}\boldsymbol{1}}{\boldsymbol{q}}_{\boldsymbol{im}} \\ 
\vdots  & \vdots  & \ddots  & \vdots  \\ 
M_{cim}{\boldsymbol{q}}^{T}_{\boldsymbol{im}}{\boldsymbol{q}}_{\boldsymbol{i}\boldsymbol{1}} & M_{cim}{\boldsymbol{q}}^{T}_{\boldsymbol{im}}{\boldsymbol{q}}_{\boldsymbol{i}\boldsymbol{2}} & \cdots  & M_{cim} \end{array}
\right]
\end{aligned}\end{equation}
\begin{equation}\label{GrindEQ__45}\begin{aligned} 
{\boldsymbol{M}}_{{\omega }_{ij}l_{ij}}=\left[ \begin{array}{cccc}
{\boldsymbol{0}}_{\boldsymbol{3}\times\boldsymbol{1}} & M_{ci1}l_{i1}{\widehat{\boldsymbol{q}}}^{2}_{\boldsymbol{i}\boldsymbol{1}}{\boldsymbol{q}}_{\boldsymbol{i}\boldsymbol{2}} & \cdots  & M_{ci1}l_{i1}{\widehat{\boldsymbol{q}}}^{2}_{\boldsymbol{i}\boldsymbol{1}}{\boldsymbol{q}}_{\boldsymbol{im}} \\ 
M_{ci2}l_{i2}{\widehat{\boldsymbol{q}}}^{2}_{\boldsymbol{i}\boldsymbol{2}}{\boldsymbol{q}}_{\boldsymbol{i}\boldsymbol{1}} & {\boldsymbol{0}}_{\boldsymbol{3}\times\boldsymbol{1}} & \cdots  & M_{ci2}l_{i2}\boldsymbol{{\hat{q}}_{i2}}^2\boldsymbol{q_{im}} \\ 
\vdots  & \vdots  & \ddots  & \vdots  \\ 
M_{cim}l_{im}{\widehat{\boldsymbol{q}}}^{2}_{\boldsymbol{im}}{\boldsymbol{q}}_{\boldsymbol{i}\boldsymbol{1}} & M_{cim}l_{im}{\widehat{\boldsymbol{q}}}^{2}_{\boldsymbol{im}}{\boldsymbol{q}}_{\boldsymbol{i}\boldsymbol{2}} & \cdots  & {\boldsymbol{0}}_{\boldsymbol{3}\times\boldsymbol{1}} \end{array}
\right]
\end{aligned}\end{equation}
\begin{equation}\label{GrindEQ__46}\begin{aligned} 
{\boldsymbol{M}}_{l_{ij}{\omega }_{ij}}=\left[ \begin{array}{cccc}
{\boldsymbol{0}}_{\boldsymbol{1}\times\boldsymbol{3}} & M_{ci1}l_{i2}{\boldsymbol{q}}^{T}_{\boldsymbol{i}\boldsymbol{1}}{\widehat{\boldsymbol{q}}}_{\boldsymbol{i}\boldsymbol{2}} & \cdots  & M_{ci1}l_{im}{\boldsymbol{q}}^{T}_{\boldsymbol{i}\boldsymbol{1}}{\widehat{\boldsymbol{q}}}_{\boldsymbol{im}} \\ 
M_{ci2}l_{i1}{\boldsymbol{q}}^{T}_{\boldsymbol{i}\boldsymbol{2}}{\widehat{\boldsymbol{q}}}_{\boldsymbol{i}\boldsymbol{1}} & {\boldsymbol{0}}_{\boldsymbol{1}\times\boldsymbol{3}} & \cdots  & M_{ci2}l_{im}{{\boldsymbol{q}}^{T}_{\boldsymbol{i}\boldsymbol{2}}\widehat{\boldsymbol{q}}}_{\boldsymbol{im}} \\ 
\vdots  & \vdots  & \ddots  & \vdots  \\ 
M_{cim}l_{i1}{\boldsymbol{q}}^{T}_{\boldsymbol{im}}{\widehat{\boldsymbol{q}}}_{\boldsymbol{i}\boldsymbol{1}} & M_{cim}l_{i2}{\boldsymbol{q}}^{T}_{\boldsymbol{im}}{\widehat{\boldsymbol{q}}}_{\boldsymbol{i}\boldsymbol{2}} & \cdots  & {\boldsymbol{0}}_{\boldsymbol{1}\times\boldsymbol{3}} \end{array}
\right]. 
\end{aligned}\end{equation}
Moreover, $\boldsymbol{C}$ is obtained as
\begin{equation}\label{GrindEQ__47}\begin{aligned} 
\boldsymbol{C}={\left[ \begin{array}{cccccccc}
{\boldsymbol{C}}_{r_l} & {\boldsymbol{C}}_{{\eta }_l} & {{\boldsymbol{C}}_{{\omega }_{1j}}}_{\ } & \cdots  & {\boldsymbol{C}}_{{\omega }_{nj}} & {\boldsymbol{C}}_{l_{1j}} & \cdots  & {\boldsymbol{C}}_{l_{nj}} \end{array}
\right]}^T
\end{aligned}\end{equation}
where the sub-matrices are
\begin{equation}\label{GrindEQ__48}\begin{aligned} 
{\boldsymbol{C}}_{r_l}=&\sum^n_{i=1}{\sum^{n_i}_{j=1}{M_{cij}\left(l_{ij}\boldsymbol{\parallel }\boldsymbol{\omega_{ij}}{\boldsymbol{\parallel }}^{2}\boldsymbol{q_{ij}}-2{\dot{l}}_{ij}\boldsymbol{\dot{q}_{ij}}\right)}} -M_Tg \boldsymbol{e_3}+\sum^n_{i=1}{M_{qi}\boldsymbol{R_l}{\widehat{\boldsymbol\Omega}}^{2}_{\boldsymbol{l}}\boldsymbol{d_i}}
\end{aligned}\end{equation}
\begin{equation}\label{GrindEQ__49}\begin{aligned} 
{\boldsymbol{C}}_{{\eta }_l}= &{\widehat{\boldsymbol\Omega}}_{\boldsymbol{l}}\left({\boldsymbol{J}}_{\boldsymbol{l}} - \sum^n_{i=1}{M_{qi}}{\widehat{\boldsymbol{d}}}^{2}_{\boldsymbol{i}}\right){\boldsymbol\Omega}_{\boldsymbol{l}}+\sum^n_{i=1}{\sum^{n_i}_{j=1}{M_{cij}{\widehat{\boldsymbol{d}}}_{\boldsymbol{i}}{\boldsymbol{R}}^{T}_{\boldsymbol{l}}\left(l_{ij}\boldsymbol{\parallel }\boldsymbol{\omega_{ij}}{\boldsymbol{\parallel }}^{2}\boldsymbol{q_{ij}} - 2{\dot{l}}_{ij}\boldsymbol{\dot{q}_{ij}}\right)}}- \sum^n_{i=1}{M_{qi}}g{\widehat{\boldsymbol{d}}}_{\boldsymbol{i}}{\boldsymbol{R}}^{T}_{\boldsymbol{l}} \boldsymbol{e_3}
\end{aligned}\end{equation}
\begin{equation}\label{GrindEQ__50}\begin{aligned} 
\boldsymbol{C}_{\omega_{ij}}= &\, M_{cij}l^2_{ij}\parallel \boldsymbol{\omega_{ij}}{\boldsymbol{\parallel }}^{2}\boldsymbol{q_{ij}}-\sum^{n_i}_{k=1 \atop k\neq j}{M_{cij}l_{ij}l_{ik}{\widehat{\boldsymbol{q}}}^{2}_{\boldsymbol{ij}}\boldsymbol{\parallel }{\boldsymbol{\omega }}_{\boldsymbol{ik}}{\boldsymbol{\parallel }}^{2}{\boldsymbol{q}}_{\boldsymbol{ik}}}  -M_{cij}l^2_{ij}\parallel \boldsymbol{\dot{q}_{ij}}{\boldsymbol{\parallel }}^{2}\boldsymbol{q_{ij}}\\&+2\sum^{n_i}_{k=1}{M_{cij}l_{ij}{\dot{l}}_{ik}{\widehat{\boldsymbol{q}}}^{2}_{\boldsymbol{ij}}{\boldsymbol{\dot{q}}}_{\boldsymbol{ik}}}-M_{cij}{\dot{l}}^2_{ij}{\widehat{\boldsymbol{q}}}^{2}_{\boldsymbol{ij}}\boldsymbol{q_{ij}} -M_{cij}l_{ij}{\widehat{\boldsymbol{q}}}^{2}_{\boldsymbol{ij}}\boldsymbol{R_l}{\widehat{\boldsymbol\Omega}}^{2}_{\boldsymbol{l}}\boldsymbol{d_i}+M_{cij}gl_{ij}{\widehat{\boldsymbol{q}}}^{2}_{\boldsymbol{ij}} \boldsymbol{e_3}
\end{aligned}\end{equation}
\begin{equation}\label{GrindEQ__51}\begin{aligned} 
{{\boldsymbol{C}}_l}_{ij}=&\sum^{n_i}_{k=1 \atop k\neq j }{M_{cij}{\boldsymbol{q}}^{T}_{\boldsymbol{ij}}\left(l_{ik}\parallel {\boldsymbol{\omega }}_{\boldsymbol{ik}}{\boldsymbol{\parallel }}^{2}{\boldsymbol{q}}_{\boldsymbol{ik}} - 2{\dot{l}}_{ik}{\boldsymbol{q}}_{\boldsymbol{ik}}\right)}-M_{cij}l_{ij}{\boldsymbol{\dot{q}}}^{T}_{\boldsymbol{ij}}\boldsymbol{\dot{q}_{ij}} +M_{cij}g{\boldsymbol{q}}^{T}_{\boldsymbol{ij}} \boldsymbol{e_3} -M_{cij}{\boldsymbol{q}}^{T}_{\boldsymbol{ij}}\boldsymbol{R_l}{\widehat{\boldsymbol\Omega}}^{2}_{\boldsymbol{l}}\boldsymbol{d_i}\\
&+K_{ij}\left(l_{ij}-L_{ij}\right)\frac{1+\sign\left(l_{ij}-L_{ij}\right)}{2} +b_{ij}{\dot{l}}_{ij}\frac{1+\sign\left(l_{ij}-L_{ij}\right)}{2} .
\end{aligned}\end{equation}

Finally, the control input is defined by 
\begin{equation}\label{GrindEQ__52}\begin{aligned} 
\boldsymbol{P}={\left[ \begin{array}{cccccccc}
{\boldsymbol{P}}_{r_l} & {\boldsymbol{P}}_{{\eta }_l} & {{\boldsymbol{P}}_{{\omega }_{1j}}}_{\ } & \cdots  & {\boldsymbol{P}}_{{\omega }_{nj}} & {\boldsymbol{P}}_{l_{1j}} & \cdots  & {\boldsymbol{P}}_{l_{nj}} \end{array}
\right]}^T 
\end{aligned}\end{equation}
where\begin{equation}\label{GrindEQ__53}\begin{aligned} 
{\boldsymbol{P}}_{r_l}=-\sum^n_{i=1}{f_i\boldsymbol{R_i} \boldsymbol{e_3}}
\end{aligned}\end{equation}
\begin{equation}\label{GrindEQ__54}\begin{aligned} 
{\boldsymbol{P}}_{{\eta }_l}=-\sum^n_{i=1}{{f_i{\widehat{\boldsymbol{d}}}_{\boldsymbol{i}}{\boldsymbol{R}}^{T}_{\boldsymbol{l}}\boldsymbol{R_i}\boldsymbol{{e}}}_3}\ 
\end{aligned}\end{equation}
\begin{equation}\label{GrindEQ__55}\begin{aligned} 
{\boldsymbol{P}}_{{\omega }_{nj}}=l_{ij}{f_i{\widehat{\boldsymbol{q}}}^{2}_{\boldsymbol{ij}}\boldsymbol{R_i}\boldsymbol{{e_3}}}
\end{aligned}\end{equation}
\begin{equation}\label{GrindEQ__56}\begin{aligned}
{\boldsymbol{P}}_{l_{ij}}={f_i{\boldsymbol{q}}^{T}_{\boldsymbol{ij}}\boldsymbol{R_i}\boldsymbol{{e_3}}}.
\end{aligned}\end{equation}

In equation \ref{GrindEQ__52}, the components represent the generalized forces in the context of Lagrange equations. These forces are vital because they capture the effects of various factors on the system dynamics. Specifically, in the case of a quadrotor, these generalized forces encapsulate the impact of thrust generated by the quadrotor on the load it's carrying, as well as the interactions with the components of the spring-mass-damper system. Understanding these forces is crucial because they provide insights into how the quadrotor's propulsion and interaction with its surroundings affect the overall behavior and stability of the system.

\bibliography{sample}

\end{document}